\documentclass{pasj02}
\usepackage{amsmath}

\let\leftrightarrows\undefined
\let\gtrsim\undefined

\let\lesssim\undefined

\let\varnothing\undefined
\usepackage{amssymb}

\usepackage{url}
\usepackage{ulem}
\usepackage{placeins}
\usepackage{comment}
\usepackage{caption}
\Received{$\langle$reception date$\rangle$}
\Accepted{$\langle$acception date$\rangle$}
\Published{$\langle$publication date$\rangle$}

\newcommand{\ergps}{erg~s$^{-1}$}

\def\xmm{{\it XMM-Newton }}

\def\xmm{{\it XMM-Newton}}

\def\xrism{{\it XRISM}}

\def\apj{ApJ}
\def\mnras{MNRAS}
\def\aap{A\&A}
\def\apjl{ApJ}
\def\apjs{ApJS}
\def\araa{ARA\&A}
\def\pasj{PASJ}
\def\nat　{Nature}
\def\iaucirc{IAU~circular}
\def\procspie{Proc. SPIE}
\def\gca{GeCoA}
\def\ssr{SSRv}

\def\axj{AX~J1745.6$-$2901}
\def\sgra{Sgr~A}
\def\sgras{Sgr~A$^{\star}$}
\def\sgraeast{Sgr~A East}

\def\Fevc{Fe~{\sc xxv}}
\def\Fevs{Fe~{\sc xxvi}}

\def\Nivs{Ni~{\sc xxviii}}

\def\resolve{Resolve}
\def\xtend{Xtend}

\def\kms{km s$^{-1}$}

\begin{document}

\title{Narrow iron- and nickel-K absorption lines from the eclipsing low-mass X-ray binary {\axj} }
\author{Kojiro Tanaka\altaffilmark{1}}
\author{Yoshitomo Maeda\altaffilmark{2}}

\author{Ryota Tomaru\altaffilmark{3}}

\author{Lia Corrales\altaffilmark{4}}
\author{Mar{\'i}a D{\'i}az Trigo\altaffilmark{5}}
\author{Chris Done\altaffilmark{6}}
\author{Tadayasu Dotani\altaffilmark{2}}

\author{Manabu Ishida\altaffilmark{2}}
\author{Satoru Katsuda\altaffilmark{7}}
\author{Yoshiaki Kanemaru\altaffilmark{2}}
\author{Richard Kelley\altaffilmark{8}}
\author{Aya Kubota\altaffilmark{9}}
\author{Hironori Matsumoto\altaffilmark{3}}

\author{Masayoshi Nobukawa\altaffilmark{10}}

\author{Megumi Shidatsu\altaffilmark{11}}
\author{Randall Smith\altaffilmark{12}}
\author{Hiromasa Suzuki\altaffilmark{2}\altemailmark}

\author{Hiromitsu Takahashi\altaffilmark{13}}
\author{Yohko Tsuboi\altaffilmark{14}}
\author{Hideki Uchiyama\altaffilmark{15}}

\author{Shigeo Yamauchi\altaffilmark{16}}
\author{Anje Yoshimoto\altaffilmark{16}}
\author{Q. Daniel Wang\altaffilmark{17}}

\author{Jon M.\ Miller\altaffilmark{4}}
\author{Frederick S.\ Porter\altaffilmark{8}}
\author{Shinya Yamada\altaffilmark{18}}

\altaffiltext{1}{Department of Physics, Tokyo Metropolitan University, Tokyo 192-0397, Japan} 
\altaffiltext{2}{Institute of Space and Astronautical Science (ISAS), Japan Aerospace Exploration Agency (JAXA), Kanagawa 252-5210, Japan} 
\altaffiltext{3}{Department of Earth and Space Science, Osaka University, Osaka 560-0043, Japan} 
\altaffiltext{4}{Department of Astronomy, University of Michigan, MI 48109, USA} 
\altaffiltext{5}{ESO, Karl-Schwarzschild-Strasse 2, 85748, Garching bei München, Germany}
\altaffiltext{6}{Centre for Extragalactic Astronomy, Department of Physics, University of Durham, South Road, Durham DH1 3LE, UK}
\altaffiltext{7}{Department of Physics, Saitama University, Saitama 338-8570, Japan} 
\altaffiltext{8}{NASA / Goddard Space Flight Center, Greenbelt, MD 20771, USA}
\altaffiltext{9}{Department of Electronic Information Systems, Shibaura Institute of Technology, Saitama 337-8570, Japan} 
\altaffiltext{10}{Department of Teacher Training and School Education, Nara University of Education, Nara 630-8528, Japan} 
\altaffiltext{11}{Department of Physics, Ehime University, Ehime 790-8577, Japan} 
\altaffiltext{12}{Center for Astrophysics | Harvard \& Smithsonian, 60 Garden Street, Cambridge, MA 02138}
\altaffiltext{13}{Department of Physics, Hiroshima University, Hiroshima 739-8526, Japan} 
\altaffiltext{14}{Department of Physics, Chuo University, Tokyo 112-8551, Japan} 
\altaffiltext{15}{Faculty of Education, Shizuoka University, Shizuoka 422-8529, Japan} 
\altaffiltext{16}{Department of Physics, Nara Women's University, Nara 630-8506, Japan} 
\altaffiltext{17}{Department of Astronomy, University of Massachussets Amherst, 710 North Pleasant Street Amherst, MA 01003, USA}
\altaffiltext{18}{Department of Physics, Rikkyo University, Tokyo 171-8501, Japan}

\email{maeda.yoshitomo@jaxa.jp and tanaka-kojiro@ed.tmu.ac.jp (ver 2025.10.15} 


\maketitle

\begin{abstract}

We report the presence of a highly ionized absorber in the transient,
eclipsing low-mass X-ray binary \axj, observed from Feb.~26 to 29, 2024
with \xrism's \resolve\ and \xtend\ instruments. During a soft/high state
without dips, \resolve's high spectral resolution
($E/\Delta E \sim 1000$, full width at half maximum) revealed narrow
velocity widths ($\sigma \sim 110~{\rm km~s^{-1}}$) for Fe~{\sc xxvi}
and Ni~{\sc xxviii} lines, even with low photon statistics.
These widths are consistent with binary orbital motion.

The observed modest blueshift velocity
($\sim160~{\rm km~s^{-1}}$) indicates that the absorber is located
sufficiently far from the neutron star ($>10^9$~cm), so that
gravitational redshift effects are not dominant.
On the other hand, broad-band spectral analysis using a photoionized
plasma model applied to the \xtend\ data constrains the absorber to lie
within a radius of $\lesssim10^{9.5}$~cm, as inferred from the upper
limits of the best-fit ionization parameter (log~$\xi \sim 4.4$) and
the large column density ($\sim1.6\times10^{24}~{\rm cm^{-2}}$).
At this distance, the observed outward velocity of the absorber is
about an order of magnitude smaller than the escape velocity from
the neutron star.

\end{abstract}

\section{introduction}

%


Accretion disk wind and accretion disk atmosphere play essential roles in the dynamics and evolution of accretion disks. An accretion disk wind serves as an effective mechanism for angular momentum transport, facilitating mass accretion onto the central object \citep{Proga2002}, while strong winds can also lead to significant mass loss, thereby regulating the accretion rate \citep{Ponti2012}. Furthermore, the presence of disk winds alters the thermal structure and stability of the accretion disk, whereas the accretion disk atmosphere contributes to radiation reprocessing and spectral formation \citep{Done2007}. 

High-resolution spectroscopy is a powerful tool for studying accretion disk wind and accretion disk atmosphere by analyzing their velocity structures, absorption line profiles, and ionization states. Accretion disk wind is characterized by highly blueshifted absorption lines, often reaching velocities of hundreds to thousands of \kms, due to its strong outflowing motion, whereas the accretion disk atmosphere remains nearly static, exhibiting absorption lines close to the systemic velocity \citep{Proga2002}. The presence of a P Cygni profile, consisting of a blueshifted absorption component and a redshifted emission component, serves as a strong spectral signature of disk winds, while the accretion disk atmosphere generally produces narrow absorption lines without such a feature \citep{Done2007}. 

Observations using high resolution spectroscopic instruments in the X-ray regime,  enable precise measurements of these properties, allowing for a comprehensive understanding of the role of accretion disk winds and atmospheres in accretion processes.



%

A low-mass X-ray binary \axj\ was serendipitously detected in the field of view of the Galactic Center and identified as a transient X-ray burster during 1993--1994 ASCA observations (\cite{Maeda1996}; \cite{Kennea1996}). Eclipses with a period of $8.356\pm0.008$ hours suggest that \axj\ is a very high-inclination, low-mass X-ray binary containing a neutron star. \citet{Maeda1996} reported excess soft X-rays during eclipses, attributed to scattering by interstellar dust.

\axj\ is a variable source characterized by distinct quiescent and bright flux states (\cite{Maeda1996}; \cite{Muno2003}; \cite{Ponti2015}). The bright state is further divided into soft and hard states based on spectral shape, with the soft state exhibiting higher flux and the hard state associated with lower flux. Re-brightening of \axj\  was reported in June, 2023 during the Swift monitoring (\cite{Reynolds2023}) about 8 month before our observations. 

Spectral studies of these states have been conducted using various X-ray observatories. Using Suzaku, \citet{Hyodo2009} identified dips and Fe~K absorption lines (including \Fevc\ and \Fevs\ K$\alpha$ and K$\beta$) during the soft state, suggesting an outflowing absorbing gas with a bulk velocity of $\sim10^3$~km~s$^{-1}$. 

\citet{Ponti2015} conducted detailed X-ray analyses of \axj\ using 38 \xmm\ observations, including 11 capturing the source in its bright state before 2013. \citet{Ponti2017} built the large data base including the data taken in 2014--2016 and analyze them. 
Prominent Fe~K absorption lines were observed in the soft state but disappeared in the hard state. This variability does not appear to result from changes in the ionizing continuum. The small K$\alpha$/K$\beta$ ratio of the \Fevc\ and \Fevs\ equivalent widths suggests that the absorbing plasma has a column density of $N_{\rm H} \simeq 10^{23}$~cm$^{-2}$ and turbulent velocities of $v_{\rm turb} \simeq 500$--$700$~km~s$^{-1}$ or $<1000$~km~s$^{-1}$ if we consider a systematic uncertainty of the response function.  

\citet{Trueba2022} analyzed high-energy resolution spectra obtained with \textit{Chandra} HETG. They detected K-absorption lines from iron and nickel and determined the absorber's line width to be $100^{+40}_{-30}$ \kms\ and the line shift to be $270^{+240}_{-230}$ \kms. Due to limited statistics, the confidence level of these error ranges is at the 1$\sigma$ level. 
They noted that future high-resolution X-ray missions, such as \xrism, will further refine the errors of these parameters. 

The precise position of \axj\ was determined with arcsecond accuracy by \textit{Chandra} ACIS. In its quiescent state, \axj\ is identified as CXOGC J174535.6$-$290133, with a luminosity of $L_{\rm X} \sim 10^{32}$~\ergps. It is located $87\farcs$ from \sgras\ (\cite{Muno2003}; \cite{Heinke2008}).
Due to the crowding of celestial objects in this region, a solid identification of an infrared counterpart has not been made. As a result, the proper motion of this object remains unknown. 

This paper is organized as follows: Section~2 describes the \xrism\ observations and data reduction methods. Section~3.1 provides an overview of imaging and region selection for temporal and spectral analyses. Section~3.2 shows temporal analysis using the \xtend\ dataset, while Section~3.3 present the results of \resolve's high-resolution spectroscopy for the first time, along with broadband spectral analysis performed with \xtend. We present our new findings in Section~4, and discuss them in Section~5. Our results are summarized in Section~6. Throughout this paper, fitting errors are quoted at the 90\% confidence level otherwise mentioned. The error bars of the data in plots are 1 $\sigma$ error. 
The distance to \axj\ is assumed to be the same as \sgras, at 8.277 kpc (\cite{GRAVITY2022}).  

\section{Observations and data reduction}

\xrism\ observations of the \sgra\ region were conducted from 02:22:11 on February 26, 2024, through 09:29:48 on February 29, 2024, in UTC (OBSID 300044010).
The primary purpose of the observations was to perform spectroscopy of \sgraeast\ and its surrounding diffuse emission using the \resolve\ instrument. Results related to \sgraeast\ were published in a companion paper (\cite{Sgraeast2024}). The aim point for these observations was $(\mathrm{RA}, \mathrm{Dec}) = (266.4227^{\circ}, -28.9972^{\circ})$ in J2000.

\xrism\ is equipped with two scientific instruments: \resolve\ and \xtend. \resolve\ is an X-ray microcalorimeter focal-plane detector that provides non-dispersive, high-resolution spectroscopy in the X-ray band (\cite{Ishisaki2025}). \xtend\ is an X-ray CCD detector with a wide field of view of $38^\prime \times 38^\prime$ (\cite{Hayashida2018}). After standard data reduction, the effective exposure times for \resolve\ and \xtend\ were 126~ks and 108~ks, respectively.

Observations with \resolve\ were conducted with the aperture door closed which is equipped with a $\sim$270~$\mu$m thick beryllium (Be) window (\cite{Midooka2021}), limiting the bandpass to energies above 1.7~keV. Gain and energy calibrations for the \resolve\ detector require on-orbit measurements for time-dependent gain variations. A $^{55}$Fe radioactive source mounted on the filter wheel (\cite{Shipman2025}) was intermittently rotated into the aperture to calibrate the gain (\cite{Maeda2025}). Photon energies were assigned to events using fiducial gain curves derived from ground calibrations and the standard nonlinear energy scale interpolation method (\cite{Porter2025,Sawada2025}). This approach achieved an energy resolution of $\sim$4.5~eV at 6~keV in full width at half maximum (FWHM). 
Systematic uncertainties in the energy scale is $\pm0.3$~eV in the 5.4--8.0~keV band, which is equivalent to a velocity shift of $\pm15$ \kms\  (\cite{Porter2025,Eckart2024}). 
The angular resolution is $\sim 1\farcm2$ in half-power diameter (\cite{Hayashi2024}).

\begin{figure*}
  \begin{center}
  \resolve \hspace{6cm} \xtend \\
\includegraphics[width=0.8\textwidth]{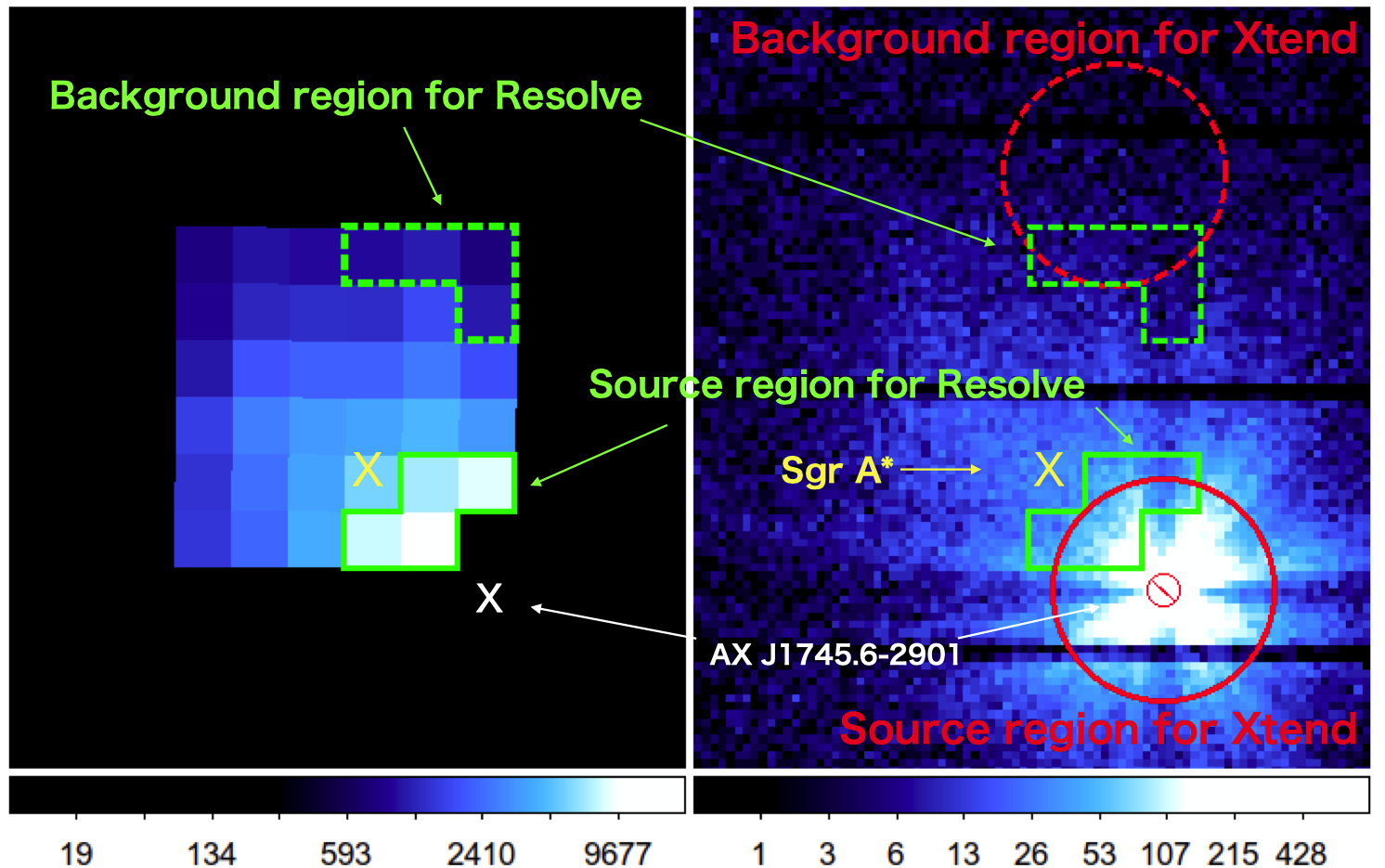}
  \end{center}
  \caption{
(Left) \resolve\ and (right) \xtend\ images in the 2–10 keV band are shown in detector coordinates, with north oriented upwards. The field of view for \resolve\ spans approximately $3' \times 3'$. The source \axj\ is located just outside the corner of the \resolve\ pixel array but remains within the field of view of \xtend. Regions used for temporal and spectral analyses are indicated with solid lines for the source ($0\farcm15$ -- $1\farcm00$ radius annual) and dashed lines for the background ($1\farcm00$~radius circle). These regions for \resolve\ are also overlaid on the \xtend\ image for comparison.}
 {Alt text: In the left image of the \resolve\ $6\times6$ array, Regions are outlined with red line. \sgras\ is within the field of view, but no significant X-ray emission has been detected from it. In the lower right, there are pixels that were not read out for calibration purposes, and \axj\ has been detected in the vicinity of those pixels. The spectrum was extracted from four pixels near that area, and the background was taken from four pixels in the upper left region. The image on the right is from \xtend. Regions are outlined with red and green lines. Since \xtend\ has a wider field of view than Resolve, \axj\ is clearly visible within the field.}     
 
  \label{fig:images}
\end{figure*}

The \xtend\ instrument (\cite{Hayashida2018}) achieves broad-band sensitivity in the 0.4--13~keV band. The CCD array (\cite{Noda2025}) was operated in full-window mode. One of the charge injection rows was set to pass across the aim point
in the early phase observations of Xtend including the one near \sgras (\cite{Suzuki2025}). As a result, the dead rows caused by charge injection do not appear at the image focus of the $1\farcm5$ off-axis source, \axj\ (see Figure~\ref{fig:images}).
The angular resolution of \xtend\ (\cite{Tamura2024}) is similar to that of \resolve.

Data reduction was performed using the pre-release Build 7 \xrism\ software with HEAsoft ver.~6.32 (\cite{Heasarc2014}) and the calibration database (CALDB) ver.~8 (v20240815) (\cite{Terada2021, Loewen2020}). Cleaned event files were generated, excluding periods affected by Earth eclipses, the sunlit Earth limb, and South Atlantic Anomaly passages.
A barycentric correction is applied only to the event arrival time, not to the event energy. The energy shift caused by Earth's orbital motion around the Sun ($-28 \, \text{km s}^{-1}$) on February 26, 2024, was not corrected in the spectral analysis but was accounted for in the table that provides the best-fit values or in the corresponding text.

For spectral analysis, we used XSPEC ver.~12.13.1 (\cite{Arnaud1996}) with the $\chi^2$ or $C$-statistic (\cite{Cash1979}) and AtomDB ver.~3.0.9 (\cite{Foster2012}). The statistical error was determined using MCMC techniques for $C$-statistic.
Redistribution matrix files (RMFs) were generated using the {\tt rslmkrmf} task, based on the cleaned event files and CALDB data from ground-based measurements. The line-spread function components include a Gaussian core, an exponential tail at lower energies, escape peaks, and Si fluorescence. Auxiliary response files (ARFs) were generated with the {\tt xaarfgen} task.

\section{Data analysis and results}

\subsection{Images}

Figure~\ref{fig:images} shows the \resolve\ and \xtend\ images around \axj. 
The \xtend\ instrument clearly detected \axj\ within its field of view. In contrast, \axj\ is positioned just outside the corner of the \resolve\ pixel array, as \sgraeast\ was placed near the center of the array. Events corresponding to the point spread function (PSF) tail from \resolve's X-ray mirror assembly (XMA) were detected from \axj.

\subsection{Lightcurves}

We made a temporal analysis using the \xtend\ data. 
We extracted a light curve from the $0\farcm15$--$1\farcm00$ radius annular region shown in Figure~\ref{fig:images}. 
The inner radius was determined based on the pixel-by-pixel grade branching ratios to exclude piled-up events.
Eclipses are clearly visible, occurring approximately every 30,000 seconds, consistent with previous observations (\cite{Ponti2017} and references therein). We detected three ingress events and four egress events clearly. 

A period search was conducted using an epoch-folding technique applied to the light curve, $c(t)$. The folded light curve was then fitted with a canonical function:

\begin{align}
c(t)\ &= c_{\rm per} + c_{\rm ecl},           & (t < t_0) \nonumber \\
                           &=  c_{\rm per} e^{-\frac{t - t_0}{\tau_0}} + c_{\rm ecl}, &(t_0 < t < t_1) \nonumber \\
                           &=  c_{\rm per} (1 - e^{-\frac{t - t_1}{\tau_1}}) + c_{\rm ecl}, &(t_1 < t)              
\end{align}

where $c_{\rm per} + c_{\rm ecl}$ is the persistent count rate outside the eclipse, $c_{\rm ecl}$ is the residual count rate during the eclipse, $t_0$ and $t_1$ are ingress and egress of the eclipse, and $\tau_0$ and $\tau_1$ are the decay and rise timescales for the ingress and egress phases, respectively.

The best-fit parameters at the 1$\sigma$ confidence level were determined as:
$t_1 - t_0 = 0.0474\pm0.0002\ {\rm phase},\ \tau_0 = (6\pm3)\times10^{-4}\ {\rm phase},\ \tau_1 = (10\pm3)\times10^{-4}\ {\rm phase},\ c_{\rm per} = 2.44\pm0.06,\ c_{\rm ecl} = 1.22\pm0.05$.
Using these best-fit parameters, we fitted each individual eclipse and determined the epochs $t_0$ for all eclipses. 

\begin{figure}[h]
  \begin{center}
    \includegraphics[bb=100 0 900 500,width=0.6\textwidth]{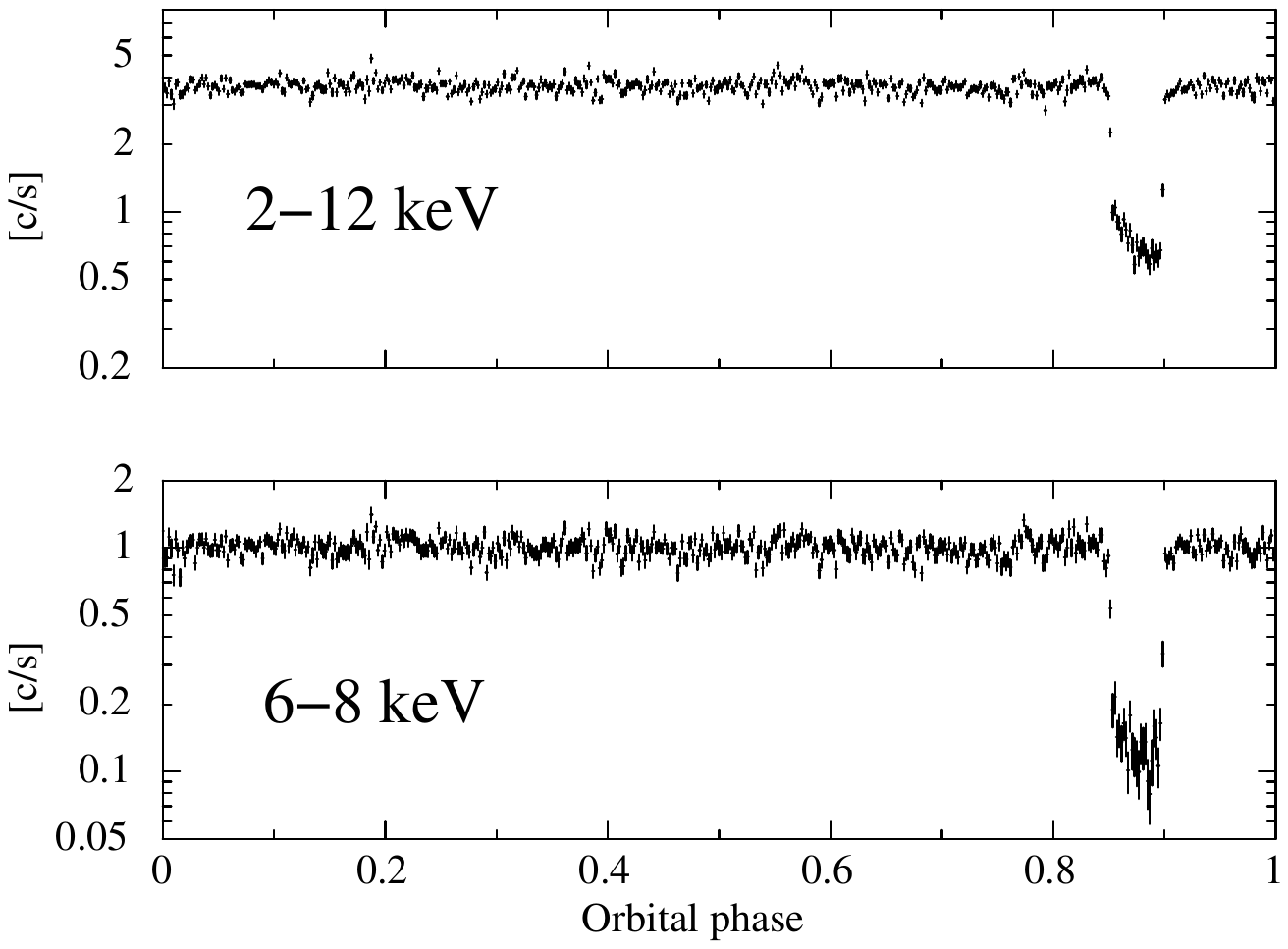}\\
  \end{center}
  \caption{
    Folded light curves of \xtend\ in the 2--10~keV (upper) and 6--8~keV (lower) bands, with phase 0 defined at MJD 60366.029. The folded period is $30,063$~s. The background was not subtracted.}
     {Alt text: The data is suddenly dropped by a factor of about five at phase 0.85 and increased at 0.90 in both plots.}
  
  \label{fig:lightcurves}
\end{figure}

By fitting $t_0$ with a linear function of a revolution, the orbital period was derived as $30,063\pm2$~s at 1$\sigma$ confidence level. The period is consistent with past measurements (e.g., $30063.6292
\pm0.0006$ at 1$\sigma$ confidence level : \cite{Ponti2017}). 
A folded light curve based on this period was created and is presented in Figure~\ref{fig:lightcurves}. The averaged count rate in the 2--10~keV band is 3.5~c~s$^{-1}$ for \xtend\ and 0.31~c~s$^{-1}$ for \resolve. The \resolve light curve shows a similar trend to \xtend\ but with significantly fewer statistics.

The eclipse fraction is approximately 0.05 phase. Regardless of energy, the light curve decays during eclipse ingress and rises with an inversion during egress both in the 2--12 and 6--8 keV band (Figure~\ref{fig:lightcurves}). As presented in previous observations (\cite{Jin2018} and references therein), these are typical features of dust scattering. Additionally, no time variability characteristic of persistent emission is observed, and the dips frequently seen in past observations (\cite{Ponti2017} and references therein) are absent.

\subsection{Spectra}


Because the number of photons in the \resolve\ spectrum is limited, we constrained only the line shift and width using \resolve.
In contrast, the broader energy coverage and higher photon statistics from \xtend\ allow us to determine the continuum shape (luminosity, absorption, and temperature). Furthermore, by incorporating the line shift and width obtained from the \resolve\ data, we were able to determine the ionization parameter from the \xtend\ spectrum, despite its limited energy resolution.

\subsubsection{ Background and energy band selection}


Backgrounds were chosen according to the specific characteristics and conditions of each instrument. For \xtend, the source spectra were created from the persistent phase, 
while the eclipse phase was used as the background. 

The widest available energy range, 2–12 keV, was adopted.
For \resolve, both the source and background spectra were generated from the persistent phase, with the background taken from a nearby region. The analysis was restricted to energies above 6.8 keV, where the systematic uncertainty introduced by subtracting the background from a different region is small.

These background regions include three major components: the Galactic diffuse emission, a leakage from the PSF tail of \sgraeast, and the dust-scattering halo (or scattered-in component), mainly originating from \axj\ itself (e.g., \cite{Maeda2002}; \cite{Predehl2003}; \cite{Jin2017}).

The main difficulty lay in subtracting the first two background components, which we addressed first. Because the same spectrum is expected during both the non-eclipse (persistent) and eclipse phases, using the eclipse spectra as the background allows these components to be almost completely removed. Therefore, this method was adopted as the primary approach.

Persistent and eclipse phases both of \xtend\ and \resolve\ were defined as 0.00--0.84/0.91--1.00 and 0.85--0.90, respectively. 
In the case of \xtend, subtracting the spectrum obtained during the eclipse phase effectively removes both the Galactic diffuse emission and the PSF leakage from \sgraeast\ over a broad energy range since a sufficient number of photons were collected during the eclipse.



%




%
In contrast, due to limited photon statistics in \resolve, we were not able to select the eclipse spectra as a background but to adopt a spectrum taken from a nearby region as a second-best option. 
A background region was selected from within the \resolve\ field of view (green dashed line in Figure~\ref{fig:images}). 

The source and background spectra are also shown in Appendix 1 for reference.

We evaluated the appropriateness of this background region using \xtend, which allows flexible selection of extraction regions.
Using \xtend, we extracted several background regions near \sgras. We then found that the \xtend\ spectra, subtracted using the circular region marked by the dashed red line in Figure~\ref{fig:images}, showed an equivalent width of $\sim$12 eV for the Ly$\alpha$ lines, consistent with the eclipse background subtraction method. Consequently, the dashed green region near the \xtend\ circular region was chosen as the background for \resolve.


As noted by previous studies (\cite{Maeda2002}; \cite{Predehl2003}), each emission line, such as Fe H-like lines, and other elements or transitions at lower energies from the Galactic diffuse emission exhibit different spatial distributions in the vicinity of \sgras. The background region that is optimal for Fe H-like lines is not necessarily optimal for other lines such as Fe He-like lines.
This introduces challenges for reasonable background subtraction in the lower energy bands (below 6.8 keV), which include emission lines from He-like Si, S, Ar, and Fe ions. To mitigate this, the \resolve\ spectral analysis was restricted to energies above 6.8 keV, excluding the lower energy range.

\subsubsection{Modeling dust scattering}
\label{sec:dustscattering}

The remaining challenge in the background subtraction is the treatment
of the dust-scattering component intrinsic to the source.
For the \xtend\ data, the residual dust-scattering contribution was
explicitly taken into account through spectral modeling.
In contrast, for \resolve, this component was neglected because of the
limited photon statistics and the relatively small scattering fraction
in the high-energy band above 6.8~keV.

The spectra of \axj\ exhibit a combination of narrow absorption lines
and absorption edges (Figures~\ref{fig:xtendpion} and \ref{fig:resolvepion}).
The absorption lines are most likely produced by highly ionized
material in the vicinity of the source, whereas the absorption edges,
such as the feature at 7.1~keV, originate from interstellar absorption
and dust scattering by non- or weakly ionized material along the line
of sight.



\begin{figure*}[h]
  \begin{center}
\includegraphics[width=0.8\textwidth]{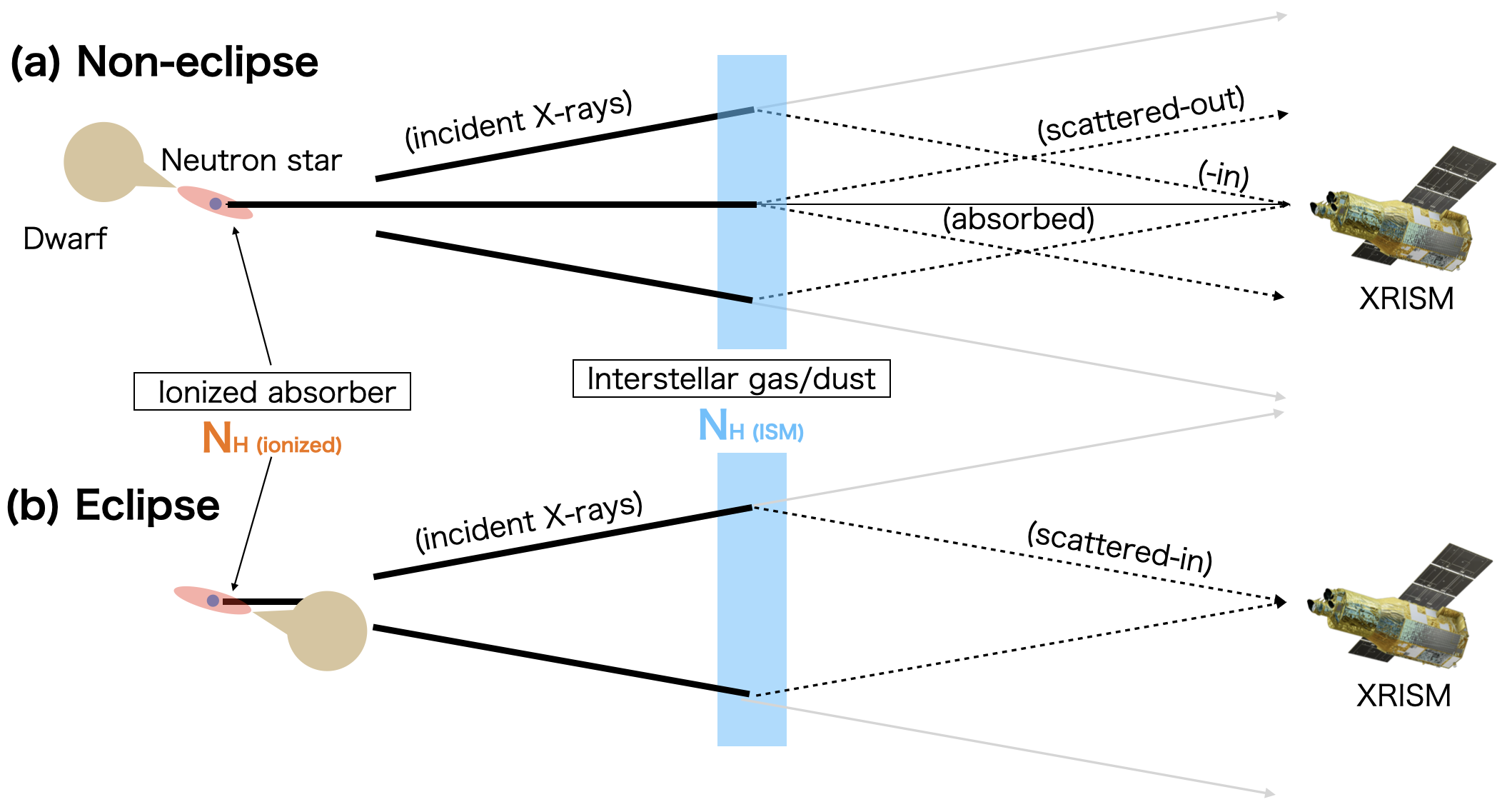}
  \end{center}
  \caption{%
A schematic view illustrating the absorption and scattering processes for \axj. Panel (a) shows the non-eclipse phase, where both direct and scattered-in light paths contribute to the observed signal. Panel (b) shows the eclipse phase, where the scattered-in component can still reach us due to the time lag between the direct light path and the scattered light path.}
  {Alt text: On the left, a binary system is shown, and on the right, the XRISM satellite is depicted. Interstellar matter lies between them. The path of the X-rays is indicated by black arrows.

}
  \label{fig:interstellar}
\end{figure*}

The top panel of Figure~\ref{fig:interstellar} illustrates absorption and scattering during the non-eclipse phase of the \axj\ binary system. X-rays emitted from the vicinity of the neutron star are absorbed by nearby structures, such as outflows (e.g., \cite{Kallman2001}), which are highly ionized by intense X-ray radiation. These X-rays then travel through interstellar space and undergo photoelectric absorption by interstellar matter (e.g., \cite{Morrison1983}) and Rayleigh-Gans scattering by dust or grains (e.g., \cite{Hayakawa1970}). Some X-rays are scattered out at grazing angles, reducing and hardening the spectrum, while others scattered in from slightly different directions are integrated into the observed photons. By accumulating photons within a sufficiently large region, most of the scattered-in photons compensate for the scattered-out photons.

The bottom panel of Figure~\ref{fig:interstellar} shows absorption and scattering during the eclipse phase. Here, emitted X-rays are completely blocked by the companion dwarf star. However, X-rays emitted before the eclipse, traveling slightly off the line of sight, may reach the observer with a time delay after being scattered toward us. As a result, the spectrum during the eclipse is
dominated by the scattered-in photons. 


\
When the intrinsic spectrum of a source varies, these variations are correspondingly delayed along the scattering paths. A complete characterization of dust scattering would therefore require long-term, continuous spectral monitoring prior to the observation, together with time-resolved imaging over a wide range of scattering angles. In practice, such information is not available, and the dust-scattering effect must be treated approximately.

\citet{ATel16436} reported that \axj\ entered an active state in July 2023, and subsequent monitoring indicated that this activity was maintained at least until 9 February 2024 (\cite{Reynolds2023}). Our observation was conducted in late February 2024, which is consistent with the source remaining active at that time. We therefore assume that the source had been active for more than one month prior to our observation and that a quasi-steady dust-scattering halo had already formed, as assumed in constructing Figure~\ref{fig:interstellar}.

\

\begin{figure}[h]
  \begin{center}
\includegraphics[width=0.49\textwidth]{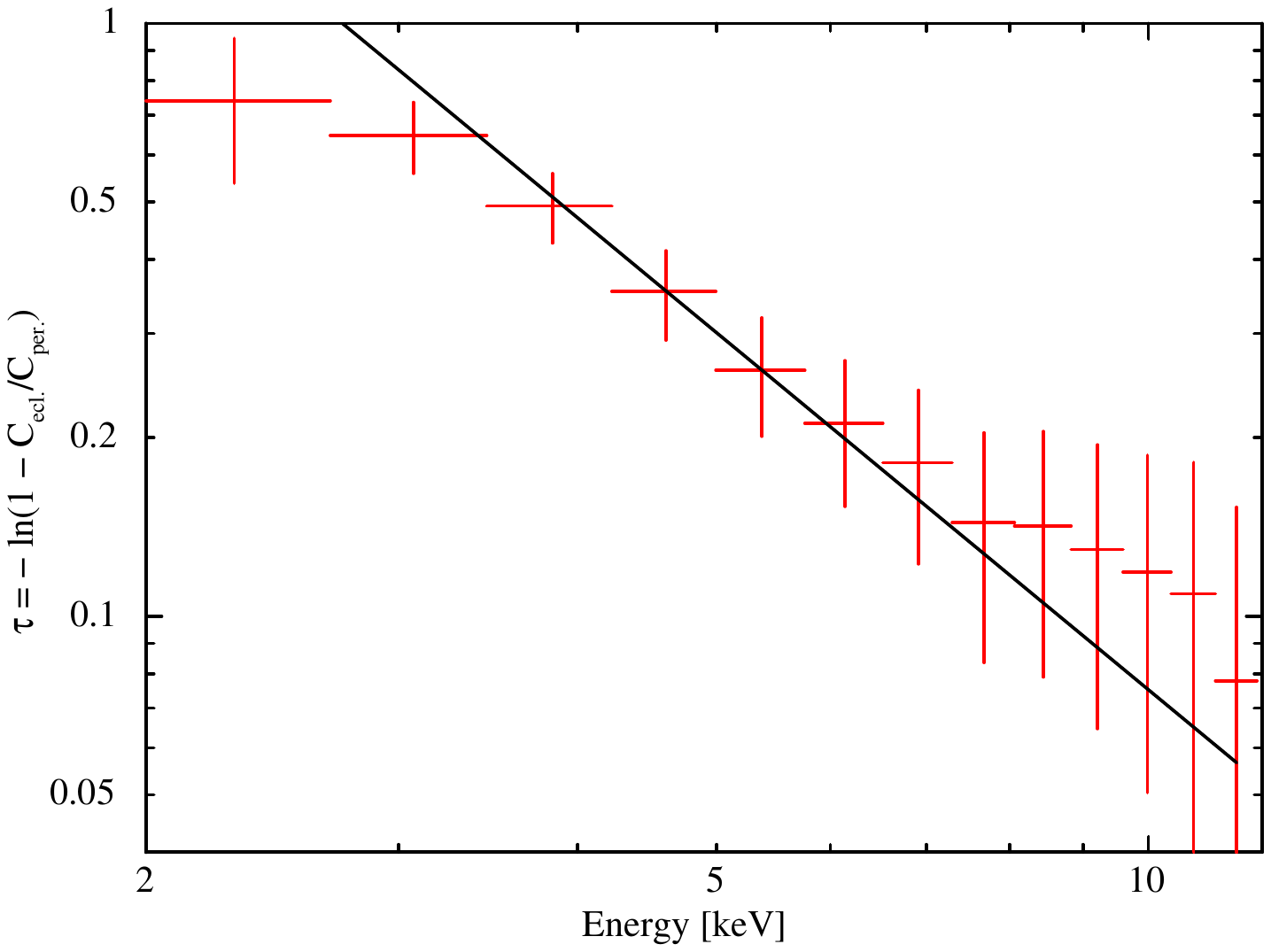}
  \end{center}
  \caption{%
The scattering cross-section of the non-eclipse to eclipse spectra, taken from the $1\farcm$ outer radius region. The solid line represents the best-fit parameters of the \(E^{-2}\) model. The excess observed above 6 keV may suggest the presence of an additional intrinsic and extended component during the eclipse. }
  {Alt text: The horizontal axis represents energy, whereas the vertical axis represents cross section. The data is shown in red.}   
  \label{fig:ratio}
\end{figure}

As long as the scattering is single, the flux ratio ${C_{\rm ecl.}}/{C_{\rm per.}} $ of the \xtend\ light curve
can be expressed as \( 1 - \exp(-\tau) \), with 
\(\tau \propto E^{-2}\) for Rayleigh-Gans scattering. We plotted \(\tau\) (i.e., \( \ln(1 - C_{\rm ecl.} / C_{\rm per.}) \)) against energy \(E\) and fitted the data with an \(E^{-2}\) model (Figure~\ref{fig:ratio}).

In order to produce the plot shown in Figure~\ref{fig:interstellar}, it was necessary to subtract the background from both the persistent and eclipse spectra. The background spectra were extracted from the red dashed region shown in Figure~\ref{fig:images}. However, since the eclipse-phase spectrum is relatively soft and has low intensity in the hard X-ray band, the choice of background region can introduce significant systematic uncertainties in that energy range. The magnitude of this uncertainty, in terms of optical depth, is approximately 0.05. This systematic uncertainty has been included in the error bars shown in Figure~\ref{fig:images}.


The \(\tau\) values in the 3--6 keV band approximately follow the \(E^{-2}\) law, consistent with the energy dependence expected for Rayleigh-Gans dust scattering. This energy-dependent behavior supports our interpretation of low-energy scattering processes, as described in Figure~\ref{fig:interstellar}.  
\citet{Mathis1991} found that for low $\tau$, multiple scatterings are approximately $\tau/2$, while at $\tau = 1$, the multiple scattering fraction is about 0.72 of the single scattering. The discrepancy from the $E^{-2}$ dependence below 3 keV may arise from neglecting the effect of multiple scattering.
%


\citet{Smith2002} found that the extended component fraction as a function of energy is approximately $1.5 \times (N_{\rm H} / 2.9 \times 10^{22}~\text{cm}^{-2}) \left(E / 1~\text{keV}\right)^{-2}$, assuming single scattering for GX~$13+1$. If we adopt $N_{\rm H} = 17 \times 10^{22}~\text{cm}^{-2}$ for \axj\ (see Table~\ref{tb:Pion}), the fraction becomes 0.35, or $\tau = 0.43$ at 5~keV. The optical depth $\tau$ measured using $C_{\rm ecl.} / C_{\rm per.}$ is 0.3. This slightly smaller value of $\tau$ suggests that part of the line-of-sight $N_{\rm H}$ is localized near the Galactic Center. Indeed, \citet{Hsieh2021} obtained the CS molecular map around \sgra, and at the south-west edge of which \axj\ is located. Around the line-of-sight velocity of $-20$ \kms\ (blueshifted), the CS molecular enhancement appears in the direction of \axj. It is possible that interstellar material, such as the Circumnuclear Disk (molecular clouds), exists locally to and in front of \axj.

Local scattering does create a small halo, resulting in a rapid decay of the scattered-in flux during the eclipse, as seen in Figure~\ref{fig:lightcurves} (see also \cite{Jin2017,Jin2018} for details). The later phase during the eclipse misses the flux of the local scattering. Since the eclipse spectra are averaged over the duration of the eclipse, the flux scattered-in from the local region is almost entirely accumulated in the persistent spectra but only partially in the eclipse spectra.
For the analysis of the \xtend\ spectra, we used the persistent phase spectra from the non-eclipse phase and subtracted the eclipse spectra. Therefore, the background of the scattered-in component is slightly under-subtracted for the \xtend\ spectra.

\

To assess its impact, we extracted spectra using apertures with radii of 1~arcmin and 5~arcmin (see Section~\ref{sec:xtendpion} for details). Although the resulting spectral parameters show slight differences, the quantitative impact is small. The absence of a significant dependence on the extraction radius suggests that rapidly varying dust-scattering components with small spatial extent have only a limited effect on the spectral analysis, implying that the contribution from a compact halo component with a characteristic timescale shorter than the eclipse duration is minor.

Empirically, the characteristic halo radius scales as $\sim$10~arcmin$/E(\mathrm{keV})$ \citep{Predehl1995}, corresponding to a time delay of $\sim$1000~s per 1~arcmin for a source at a distance of 8~kpc. At low energies, where dust scattering is more significant, the halo can extend to several arcminutes, resulting in time delays much longer than the eclipse duration.


As a consequence of these considerations,
\
interstellar absorption was modeled using the \texttt{TBabs} code (\cite{Wilms2000}) in the XSPEC fitting package, and scattering was modeled with the \texttt{xscat} code (\cite{Smith2016}) in the \texttt{xspec} package. To account for the scattered-out component only, a small extraction radius (\(R_{\rm ext} \ll 1\farcs\)) was assumed for the \texttt{xscat} modeling.




Since the background was extracted from a nearby region, shown as the green dashed area in Figure~\ref{fig:images},
the \resolve\ spectra are affected by both scattered-in and scattered-out components. However, as shown in Figure~\ref{fig:ratio}, an energy band above 6.8~keV, used for spectral fitting, is influenced by scattering up to 20\%. 
\
Therefore, the effect of dust scattering on the derived spectral parameters in this energy band is expected to be limited and was not explicitly modeled.
\



\subsubsection{Line width and shift using \resolve}

\begin{figure}[h]
  \begin{center}
  \begin{center}
                \includegraphics[width=0.55\textwidth,angle=0]{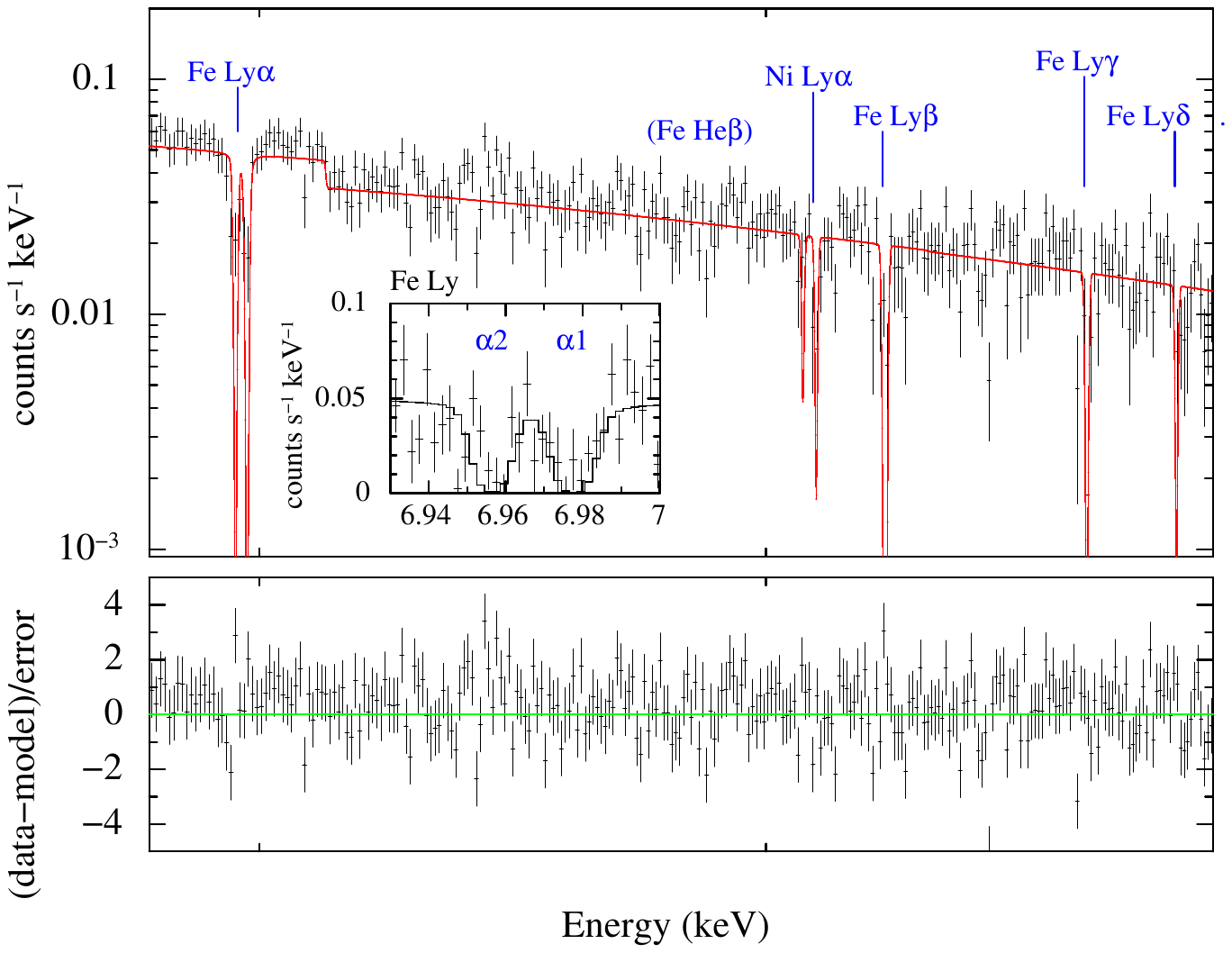}
    \end{center}
  \end{center}
  \caption{%
\resolve\ spectra in the 6.8--9~keV band fitted with the \texttt{ionabs} model. The spectra were binned for clarity in display.}
  {
  Alt text: The horizontal axis represents energy, while the vertical axis shows intensity in units of \(\mathrm{c/s/keV}\) in the upper panel, and residuals in units of \((\mathrm{data} - \mathrm{model})/\mathrm{error}\) in the lower panel. The data and residuals are shown in black, while the model is shown in red.}
     
  \label{fig:gauss}
\end{figure}

Figure~\ref{fig:gauss} shows the background-subtracted \resolve\ spectra in the 6.8--9~keV band. 
Two transitions, Ly-${\alpha1}$ and Ly-${\alpha2}$, appear separately in the \resolve\ spectra, indicating that the line widths are narrow.

We modeled the spectra using the absorption line model \texttt{ionabs}\footnote{Code can be downloaded at https://github.com/ryotatomaru/Ionabs. See details Tomaru et al. 2025 in prep.}, 
which characterizes absorption lines from a given ion in terms of the ionic column density $N_{\rm ion}$, energy shift $z_{\rm (ionabs)}$, and turbulent velocity $V_{\rm turb\ (ionabs)}$. This model calculates multiple absorption line transitions from a principal quantum number $n = 1$ for H- and He-like ions, including the full fine structure of all lines and a self-consistent edge structure. 
 So, the \texttt{ionabs} model is a useful method for deriving these characteristics  of each ions from data like this, where the photon statistics are too limited to fit individual absorption lines separately.


Given the limited photon statistics, the spectral shape of the continuum emission was fixed to the best-fit model obtained from the \xtend\ spectral analysis (Figure~\ref{fig:xtendpion}), while its normalization was left as a free parameter to account for potential effective area calibration uncertainties between the \xtend\ and \resolve\ instruments.  
The best-fit parameters are summarized in Table~\ref{tb:ionabs}. The line center exhibits a redshift of \( z = -5.4^{+1.1}_{-1.3} \times 10^{-4} \), corresponding to a blueshift of \( 160^{+50}_{-70} \)~km~s\(^{-1} \), and the line width is constrained to be as narrow as \( 110^{+40}_{-30} \)~km~s\(^{-1} \).  
Thanks to the high spectral resolution of \resolve, this narrow line width was measured for the first time.

Due to the limited photon statistics, the line width and energy shift of the nickel lines were tied to those of the irons.
\
The column density ratio of \Nivs\ to \Fevs\ was assumed to follow the solar elemental abundance ratio of Ni to Fe given by \citet{Wilms2000}.
\

\begin{table*}[h]
 \begin{center}
  \caption{\texttt{Ionabs} models for absorption lines above 6.8 keV (\resolve)}
 \label{tb:ionabs}
 \begin{tabular}{lccccc} 
 \hline \hline
  \multicolumn{4}{l}{Model : TBabs $\times$  (Disk blackbody $+$ Blackbody) $\times$ \texttt{ionabs} } \\ \hline
    \multicolumn{4}{l}{Interstellar absorption {\it TBabs}..........} \\
 & $N_{\rm H \ (TBabs)}$ & [$10^{23}$cm$^{-2}$] & 1.70(fix)  \\
        \multicolumn{4}{l}{Disk blackbody ..........} \\
 & $kT_{\rm in\ {(DiskBB)}}$  & [keV] & 1.14 (fix) \\
 & Normalization$_{\rm \ (DiskBB)}$ & [$(R_{\rm in\ km}/D_{\rm 10\ kpc})^2 \cos\theta$] &   90 (fix) \\
    \multicolumn{4}{l}{Blackbody ..........} \\
 & $kT_{\rm (BB)}$  & [keV] & 2.0(fix) \\
& Normalization$_{\rm \ (BB)}$ & [$L_{39\ \rm erg\ s^{-1}}/D_{\rm 10\ kpc} ^2$] &  1.4$\times10^{-2}$ (fix) \\ \hline

\multicolumn{4}{l}{Absorption lines (\texttt{ionabs}) ..........} \\
Ion ID & $V_{\rm turb\ \rm (ionabs)}$ [km s$^{-1}$]  & $V_{\rm los \ (ionabs)}$ [km s$^{-1}$]$^*$ & $N_{\rm ion }$ [cm$^{-2}$]\\  
\Fevs &  $110^{+40}_{-30}$ &  ${160^{+50}_{-70}}^\dagger$  &  $2.5^{+3.0}_{-1.0}\times10^{19}$ \\ 
\Nivs  &  "  &  " &  " \\ \hline
\multicolumn{4}{r}{C-Statistic              {   4796.26   }  using 4399 bins} \\ \hline
\end{tabular}
\end{center}
We fixed the continuum "TBabs $\times$ (Disk blackbody $+$ Blackhody)"  to the best-fit parameters determined by the \xtend's \\
$^*$ line of sight (los) velocity. we define the direction of blue shift as positive. \\

$^\dagger$ Equivalent to the redshift $z_{\rm \ (ionabs)}$  of  $-5.4^{+1.1}_{-1.3} \times 10^{-4}$. 
The correction of the energy shift caused by Earth's orbital motion around the Sun ($-28 \, \text{km s}^{-1}$) was applied.\\
%
%
\end{table*}

\subsubsection{Photo-ionization modeling}
\label{sec:xtendpion}

Absorption lines from highly ionized ions indicate the presence of a photo-ionized absorber along the line of sight. To model this, we applied the photo-ionization model \texttt{pion} (\cite{Mehdipour2016}) to the \xtend\ spectra in the 2--12~keV band.

For reference, we also applied the same modeling to the \resolve\ data, as shown in Appendix 2.

We modeled the absorber using the photoionization plasma code {\tt pion} in the X-ray spectral analysis software {\sc spex} \citep[v3.08.01]{Kaastra1996}.
The relevant parameter is the ionization parameter, $\xi = L_{\rm X}/nR^2$, with $L_{\rm X}$ being the source bolometric X-ray luminosity, $n$ the electron density, and $R$ the distance from the ionizing source. 
First, we created a multiplicative table model in {\sc xspec} format using {\tt pion}, assuming that the ionizing source is the X-ray emission near the neutron star, and applied it to the spectra.
The continuum for this modeling was imported into \textsc{spex} and is well described by a disk-blackbody plus blackbody model with $kT_{\rm in} = 1.14$\,keV and $kT_{\rm BB} = 2.0$\,keV, respectively (see Tab.~\ref{tb:Pion}).
We generated 10,816 transmitted spectra from the photoionized plasma slab by varying the ionization parameter $\xi$ ($1 \leq \log(\xi/\mathrm{erg~cm~s^{-1}}) \leq 6$ in steps of 0.2), the column density $N_{\rm H}$ ($20 \leq \log (N_{\rm H}/\mathrm{cm^{-2}}) \leq 25$ in steps of 0.2), and the velocity broadening $V_{\rm rms}$ ($-5 \leq \log(V_{\rm rms}/c) \leq -2$ in steps of 0.2).
Each simulation consisted of 65,536 logarithmically spaced energy bins from 1\,eV to 1\, MeV, yielding an energy width of $\sim$1.5\,eV around 7\,keV, which is sufficient for analyzing the {\resolve} spectrum.
Although {\sc spex} can fit the photoionized plasma parameters and the continuum simultaneously using the {\sc pion} model, we fixed the continuum shape to the best-fit values obtained from the {\xtend} data.
This strategy reduced both the number of free parameters and the total computational time.

\begin{figure}
  \begin{center}
\includegraphics[width=0.55\textwidth]{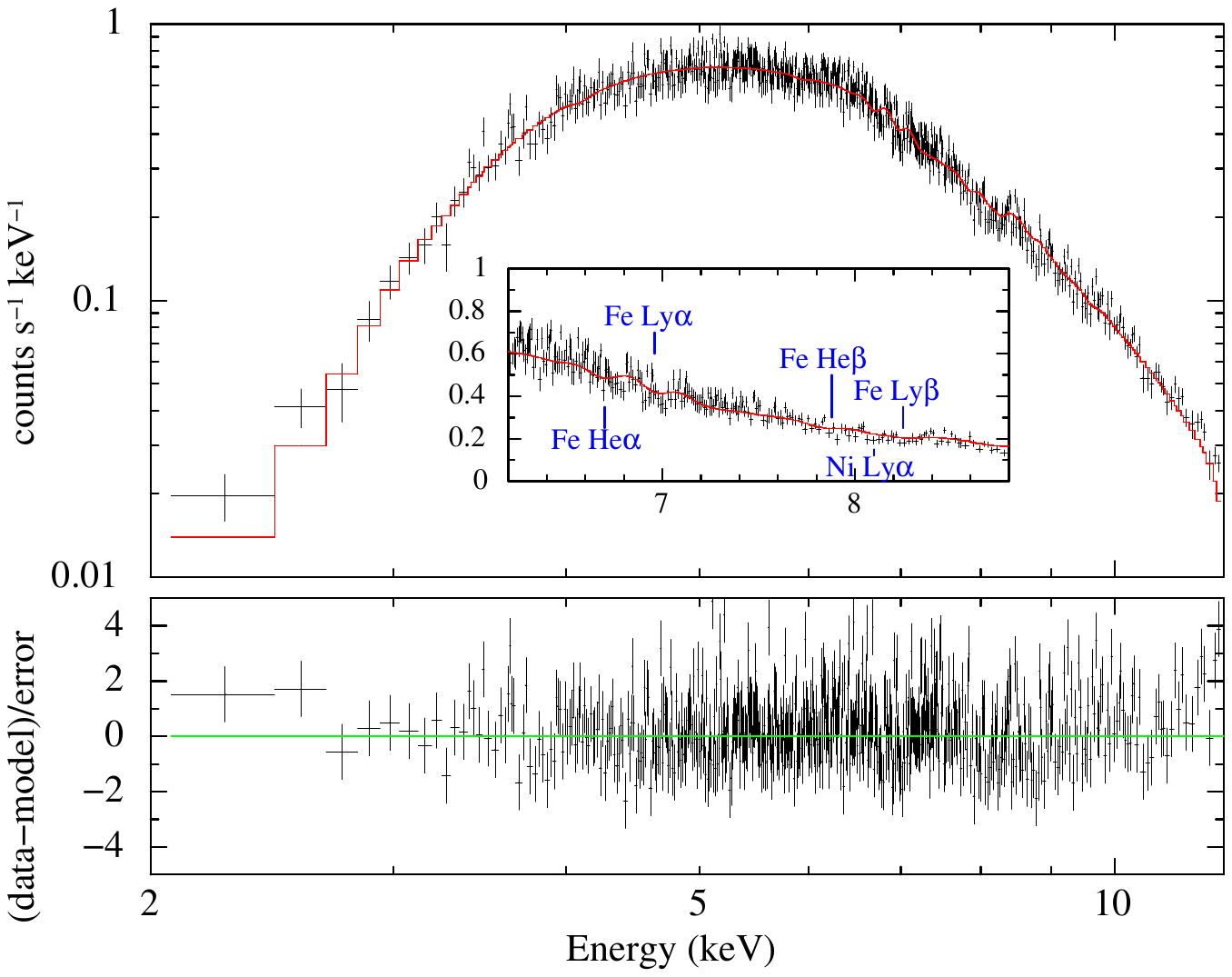}\\
  \end{center}
  \caption{%
\xtend\ spectra taken from the annular region with the $1\farcm$ outer radius. The solid line represents the best-fit \texttt{pion} model, with the residuals shown at the bottom.
}%
 {
  Alt text: The horizontal axis represents energy, while the vertical axis shows intensity in units of \(\mathrm{c/s/keV}\) in the upper panel, and residuals in units of \((\mathrm{data} - \mathrm{model})/\mathrm{error}\) in the lower panel. The data and residuals are shown in black, while the model is shown in red.}
  \label{fig:xtendpion}
\end{figure}

We fitted two spectra extracted from two annular regions with radii of $0\farcm15$--$1\farcm00$ and $0\farcm15$--$5\farcm00$.
The best-fit parameters are shown in Figure~\ref{fig:xtendpion} and listed in Table~\ref{tb:Pion}. To allow the use of $\chi^2$ statistics, each spectrum was grouped to have a minimum of 100 c per bin.

The ionization parameter $\xi$ is sensitive to the depths of the \Fevc\ and \Fevs\ absorption lines.
Observationally, it has been shown that determining the column density from an absorption feature composed of multiple blended lines, without accurately constraining the line width, can introduce uncertainties of up to an order of magnitude (e.g., \cite{Kotani2000, Kotani2006}).
To make this effect negligible, we fixed the line width to the precise value measured with \resolve.

In the Fe-K band around 7~keV, the spectrum extracted with the $1\farcm$ radius clearly exhibits a better signal-to-noise ratio (c.f., Figure~\ref{fig:Xtend_raw_bgd2}). We therefore first fitted this spectrum with the \texttt{pion} model to determine the ionization parameter \( \xi \).
However, there is a potential concern with using the $1\farcm$ extraction radius. According to Figure~5 of \citet{Jin2017}, the extent of the dust scattering halo at 5~keV is approximately $1\farcm$ or slightly less. Since the halo extent \( \theta_{\rm scat} \) scales inversely with energy (\( \theta_{\rm scat} \propto 1/E \)), it is expected to be about $2\farcm5$ at 2~keV and $1\farcm$ at 5~keV. Therefore, the $1\farcm$ extraction region may not fully capture the scattered-out component of the halo.

To assess the impact of this possible loss, we compared the spectral parameters obtained from the $1\farcm$ and $5\farcm$ extraction regions. This comparison allows us to evaluate the level of systematic uncertainty introduced by the incomplete inclusion of the scattered-out component in the $1\farcm$ spectrum. Note that in the $5\farcm$ spectrum, the source-to-background ratio in the Fe-K band is heavily contaminated from nearby bright Fe-K emitters such as \sgraeast\ 
(see Appendix 3). 
We then fixed the $\xi$ of $5\farcm$ spectrum to the best-fit values determined with the $1\farcm$ spectrum. 

Table~\ref{tb:Pion} shows that the best-fit parameters obtained from the two different regions are similar. 
\
This indicates that rapidly varying, small-angle scattering components do not significantly affect the spectral parameters derived in our analysis.
\
Based on this, we concluded that the $1\farcm$ outer radius is sufficient for spectral fitting, and we adopted the best-fit parameters from this region for further analysis.
The continuum was modeled using a disk blackbody plus blackbody. Due to significant interstellar absorption that reduces low-energy photons, the blackbody temperature was fixed to 2~keV. The spectra were well reproduced with a high ionization parameter of 

$\log \xi = 4.4^{+0.1}_{-0.2}$ 
and a thick absorber column density of $N_{\rm H (pion)} = 16^{+6}_{-7} \times 10^{23}$~cm$^{-2}$.

At a large log $\xi$ value of 4.4, the ion fractions of Fe~{\sc xxv} (He-like), Fe~{\sc xxvi} (H-like), and Fe~{\sc xxvii} (fully ionized) are approximately 1:3:6. 

In the \resolve\ spectrum shown in Figure~\ref{fig:gauss}, while the He-like K$\beta$ absorption line at around 7.88 keV appears shallow, the H-like K$\beta$ line at around 8.25 keV is deep and clearly visible. This indicates that H-like iron ions are more abundant than He-like iron, which is qualitatively consistent with the high ionization parameter ($\xi$) derived from the \xtend\ data.
At a log $\xi$ value of 4.4, 

about 30\% of the iron contributes to the absorption lines of Fe~{\sc xxvi} (H-like). The \texttt{ionabs} model gives a best-fit column density of Fe~{\sc xxvi} (H-like) of $2.5^{+3.0}_{-1.0} \times 10^{19}$ Fe cm$^{-2}$. Assuming a solar abundance ratio ($2.69 \times 10^{-5}$; \cite{Wilms2000}) and an ion fraction of 30\%, the column density for the absorber is estimated to be  $3^{+4}_{-1} \times 10^{24}$ H cm$^{-2}$, which is consistent with the column density of $1.6^{+0.6}_{-0.7} \times 10^{24}$ H cm$^{-2}$ derived from the photoionization model using the \xtend\ data.  

\begin{table*}[htbp]
 \begin{center}
  \caption{Broad-band fitting in the 2-12 keV range (\xtend)} 
 \label{tb:Pion}
 \begin{tabular}{llccc} 
 \hline \hline
  \multicolumn{5}{l}{Model : TBabs $\times$ xscat  $\times$ \texttt{pion} $\times$  (Disk blackbody $+$ Blackbody) } \\ \hline
\multicolumn{3}{l}{ Outer integrated radius} & $1\farcm$ & $5\farcm$ \\ \hline
  \multicolumn{5}{l}{Interstellar absorption {\it TBabs}..........} \\
  & $N_{\rm H \ (TBabs)}$ & [$10^{23}$cm$^{-2}$] & $1.70^{+0.07}_{-0.07}$ & $  1.56^{+0.10}_{-0.09}$ \\
   & $z_{\rm \ (TBabs)}$ & & \multicolumn{2}{c}{0(fix)} \\
   \multicolumn{5}{l}{Interstellar dust/grain scattering  {\it xscat} ..........} \\
  & $N_{\rm H (xscat)}$ & [$10^{23}$cm$^{-2}$] & \multicolumn{2}{c}{$=N_{\rm H \ (tbabs)}$} \\
    & Xpos &  & \multicolumn{2}{c}{$0.5$ (fix)} \\
    & $R_{\rm ext}$ & [$\farcs$] & \multicolumn{2}{c}{$10^{-4}$ (fix)} \\ 
    & $z_{\rm \ (xscat)}$ &  & \multicolumn{2}{c}{$0$ (fix)}\\
\multicolumn{5}{l}{Absorption by photoionized materials \texttt{pion} ..........} \\
  & log\ $\xi$ & [erg cm s$^{-1}$] &  $4.4^{+0.1}_{-0.2}$  & $=4.4$ (fix)  \\
  & $N_{\rm H (pion)}$  & [$10^{23}$cm$^{-2}$] &  $16^{+6}_{-7}$ &   $=16$ (fix)  \\
     & $V_{\rm rms\ \rm(pion)}$  & [km s$^{-1}$] & \multicolumn{2}{c}{ $110$ (fix)}\\
    & $z_{\rm \ (pion)}^*$ &  &  \multicolumn{2}{c}{ $-5.4\times10^{-4}$ (fix)} \\
    \multicolumn{5}{l}{Disk blackbody ..........} \\
 & $kT_{\rm in\ {(DiskBB)}}$  & [keV] &  $1.14^{+0.06}_{-0.05}$ &   $1.17^{+0.10}_{-0.08}$  \\
 & Normalization$_{\rm \ (DiskBB)}$ & [$(R_{\rm in\ km}/D_{\rm 10\ kpc})^2 \cos\theta$] &   $90^{+60}_{-50}$ &   $70^{+40}_{-30}$ \\
    \multicolumn{5}{l}{Blackbody ..........} \\
 & $kT_{\rm (BB)}$  & [keV] & 2.0(fix) & $=2.0$(fix) \\
& Normalization$_{\rm \ (BB)}$ & [$L_{39\ \rm erg\ s^{-1}}/D_{\rm 10\ kpc}^2$]  &  $1.4^{+0.7}_{-0.6}\times10^{-2}$ &  $1.4^{+0.1}_{-0.1}\times10^{-2}$  \\ \hline
\multicolumn{5}{l}{Total ..........} \\
& Absorbed flux$_{\rm \ 3-6 keV}$ & ergs s$^{-1}$ cm$^{-2}$ & $7.1 \times 10^{-11}$ & $6.9 \times 10^{-11}$ \\
& Interstellar extinction-collected bolometric $L_{\rm x}$  & ergs s$^{-1}$  &   $1.1 \times 10^{37}$ & $8.8 \times 10^{36}$\\ 
& All extinction collected bolometric $L_{\rm x}$ & ergs s$^{-1}$  &   $3.8 \times 10^{37}$ &     $3.1 \times 10^{37}$ \\ \hline
\multicolumn{3}{r}{Chi-Squared} &   1201.45 using 998 bins &   1538.69 using 1285 bins \\ \hline
\end{tabular}
\end{center}
We adopted the dust-grain size distribution from \citet{Mathis1977} for the {\it xscat}\ model. \\
$^*$ The parameters $z_{\rm (pion)} = z_{\rm (ionabs)}$ and $V_{\rm rms} = 1/\sqrt{2} \ V_{\rm turb\ (ionabs)}$ were fixed to the best-fit values determined by the \resolve\ \texttt{ionabs}\ fitting. The correction of the energy shift caused by Earth's orbital motion around the Sun ($-28 \, \text{km s}^{-1}$) was applied. The $z_{\rm \ (pion)}$ before the correction is  $-6.3\times10^{-4}$.  
\end{table*}

\section{Discussion}

\subsection{Absorber and its location}

Since the \resolve\ spectra are averaged over the orbital phase, the absorption line is expected to be broadened due to the neutron star's orbital motion. 
Using the orbital parameters from \citet{Maeda1996}, with a companion mass of $1.0 \, M_{\odot}$, a semimajor axis of $2.7 \, R_{\odot}$, and an orbital inclination angle of $70^\circ$, the projected rotational velocity $v \sin \theta$ is estimated to be $\sim150$~km~s$^{-1}$. This corresponds to an equivalent broadening of $v = 110$~km~s$^{-1}$ (rms), which is consistent with the line broadening of  $\sim110$~km~s$^{-1}$   obtained from the \resolve's \texttt{ionabs} fitting. The narrow line width can be fully explained by binary rotation, suggesting that the absorber moves stably with suppressed turbulence. 
Together with the low velocity shift of the absorption lines without a feature such as a P Cygni profile, the narrow line widths indicate the absorber's origin is more consistent with an accretion disk atmosphere than an accretion disk wind \citep{Done2007}.

The fitting with the photo-ionization model \texttt{pion} using \xtend\ indicates a high ionization parameter of log $\xi = 4.4^{+0.1}_{-0.2}$. 
Assuming that the extent of the absorber along the line of sight is equal to or less than its radial distance $R_{\rm abs}$, the ionization parameter $\xi$ is given by 

\begin{equation} \label{eq:xi2}
    R_{\rm abs} \leq \frac{L_X}{\xi N_{\rm H}},
\end{equation}
where $L_X$ is the X-ray luminosity and $N_{\rm H}$ is the column density of the absorber. Using the best-fit parameters from the \texttt{pion} model, the radial distance to the absorber can be estimated as:

\begin{align} \label{eq:R}
    R_{\rm abs} \leq  3.2 \times10^{9} \left( \frac{L_{\rm x}}{3.8 \times 10^{37} \ {\rm erg\ s^{-1}}} \right) \ {\rm cm} \nonumber  \\
    \approx 1.6 \times 10^{4} \left( \frac{L_{\rm x}}{3.8 \times 10^{37} \ {\rm erg\ s^{-1}}} \right) \ {R_{\rm g}},
\end{align}

where ${R_{\rm g}}$ is the gravitational radius ${{\rm G} M}/{{\rm c}^2}$ with  the neutron star mass $M$, the speed of light c, and  the gravitational constant G. We assumed a mass of $1.4~$M$_\odot$ for a neutron star. 
\
This estimate should be regarded as an order-of-magnitude upper limit, based on simplified assumptions such as a single-zone geometry and uniform density.
\

\begin{figure}[h]
  \begin{center}
        \includegraphics[width=0.48\textwidth,angle=0]{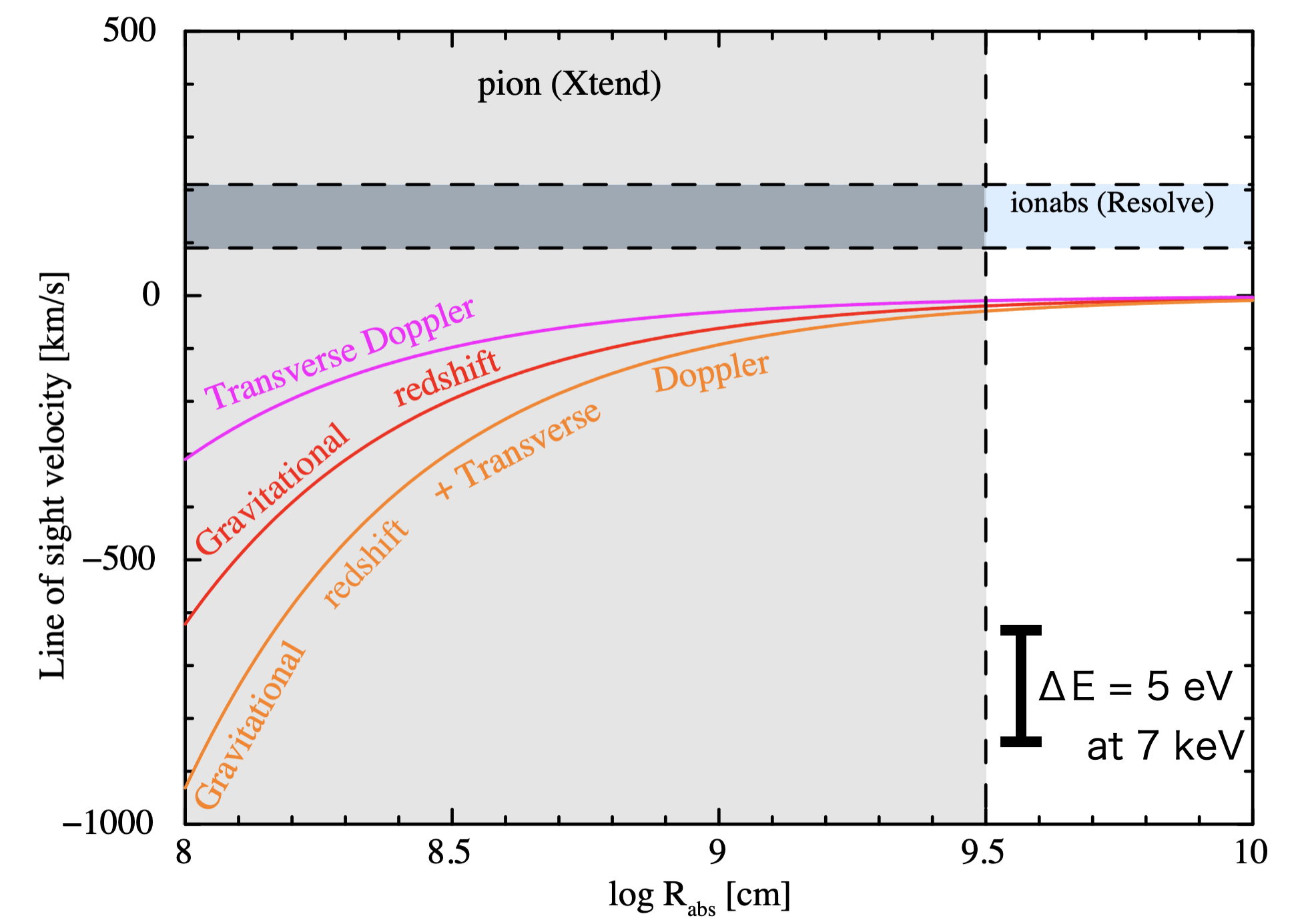}\\
  \end{center}
  \caption{%
    Plot of the radius and Doppler shift of the ionized absorbing material. The gray square represents the region allowed by our analysis, while the red and orange lines indicate the expected redshift due to the transverse Doppler effect and gravitational redshift, respectively. }
{ Alt text: The horizontal axis represents \( R_{\rm abs} \), and the vertical axis represents the line-of-sight velocity. The error region of \texttt{pion} is shown in grey, while that of \texttt{ionabs} is shown in light blue. The redshift due to the transverse Doppler effect and gravitational redshift are indicated by purple and vermilion lines, respectively. The sum of both effects is shown as an orange line.}

   \label{fig:solution}
\end{figure}

Figure~\ref{fig:solution} shows the expected location of the absorber calculated using this equation.
The absorber is determined to be located at a radius of $\leq 3.2 \times10^{9}$~cm.
The absorber's radius $R_{\rm abs}$ is at least one to two orders of magnitude smaller than the Roche lobe radius ($R_{\rm L}= 7\times10^{10}$~cm or 0.9~R$_\odot$ or $4\times10^{5}$ $R_{\rm g}$). 


At this distance, the neutron star’s strong gravitational field can induce a redshift due to the transverse Doppler effect while orbiting with Keplerian velocity, given by

\begin{equation} \label{eq:deltae} 
\frac{\Delta E}{E} =1 - \sqrt{1 - \frac{R_{\rm g}}{R}} \ \ \ \ \ \ \ ({\rm transverse\ Doppler})
\end{equation}
and a gravitational redshift, given by

\begin{equation} \label{eq:deltae2} 
\frac{\Delta E}{E} = {\frac{1}{\sqrt{1 - \frac{2 R_{\rm g}}{ R}}}}- 1 \ \ \ \ \ \ \ ({\rm gravitational\ redshift}) ,
\end{equation}
These effects become significant at distances of $10^{8-9}$~cm or closer, as shown in Figure~\ref{fig:solution}. The gravitational redshift is larger than the transverse Doppler effect. After correcting for these redshifts, we plotted the line-of-sight velocity of the absorbers in rest frame in Figure~\ref{fig:correctedvelocity}. 
The correction has only a minor effect on the velocity; at $10^{9.5}$~cm, the upper limit of $R_{\rm abs}$, the rest-frame line-of-sight velocity is almost identical to the observed value.

The rest-frame velocity of  $\sim$180~km~s$^{-1}$ ($R = 10^{9.5}$cm)  or higher may originate from either the proper motion of the binary system or an outflow from the disk. In the proper motion scenario, \axj\ would have a velocity vector directed toward us, with the absorber co-moving with the binary system. Measurements of interstellar gas in this region indicate a slow rotational velocity of only 100~km~s$^{-1}$ for a stable circular orbit \citep{Lugten1986}. The rotation velocity is similar to that of the line of sight velocity for the $10^{9.5}$ cm solution. The velocity shift may be fully explained by the proper motion, resulting in the absorber being an orbiting disk atmosphere around the neutron star without radial motion. 

Alternatively, the blueshift may result from a slow disk-driven outflow. 
The absorber may have a velocity component perpendicular to the disk plane, contributing to the observed line-of-sight velocity in \xrism. If the absorber originates from the accretion disk atmosphere, the wind speed is expected to be relatively slow ($\sim10^{1}$ \kms). This constraint implies that the absorber's location should be beyond the region significantly affected by the neutron star's gravitational redshift, i.e., at a distance of $\gtrsim 10^9$ cm (e.g., Figure~\ref{fig:solution}).  
If both the interpretation of the absorber as part of the accretion disk atmosphere and the location determined based on the ionization parameter $\xi$ are valid, the absorber's distance is estimated to be $10^9$ cm.

\citet{Hyodo2009} reported that the ionization parameter, log~$\xi$, was as low as 3.5 during the dip state, based on the fact that the absorption line of He-like Fe was the deepest. This value is significantly lower than that derived from our observations during the non-dip state (log~$\xi$ = 4.4). This discrepancy suggests that the absorber responsible for the dip, commonly referred to as the “bulge” (e.g., \cite{White1982}), is located farther out than the region probed by our observations. Assuming that the bulge lies near the Roche lobe radius ($R_{\rm L} \sim 7 \times 10^{10}$ cm), our conclusion—that the absorber detected by XRISM is not associated with the bulge but rather with the accretion disk atmosphere at a distance of $10^9$ cm or less—remains consistent.


\subsection{Comments on wind-driving mechanism}

Among wind-driving mechanisms, the most common is a thermally driven wind \citep{Begelman1983,Woods1996}. When the inner accretion disk is heated to high temperatures ( \( T \sim 10^7 \) K) due to X-ray irradiation from the neutron star, the gas expands. If its thermal energy exceeds the gravitational binding energy, it escapes as a wind. The critical radius for wind formation is approximately one-tenth of the Compton radius, where the radiation and matter become thermal equilibrium through Compton scattering. Using the best-fit parameters of \texttt{pion}, we found the Compton temperature is ($T_c \sim 1.1 \times10^7~ \rm{K}$). 
For a neutron star, the Compton radius $R_{\rm c}$ is then given by
\begin{align}
    R_{\rm c} &= 5.9\times10^5\ R_{\rm g},
\end{align}
  whereas the wind launching radius $R_0$ is 
\begin{equation}
R_0 = (0.1{\rm -}0.2)~R_{\rm c} = (6{\rm -}12)\times10^4\ R_{\rm g}.
\end{equation}
This launching radius is an order of magnitude larger than the absorber's radius, $R_{\rm abs} (\leq 1.6 \times 10^{4} , R_{\rm g})$. { If we adopt the thermally driven wind model for \axj\ in the high/soft state, the observed absorber cannot be explained as a thermally driven wind.



\begin{figure}[h]
  \begin{center}
     \includegraphics[width=0.48\textwidth]{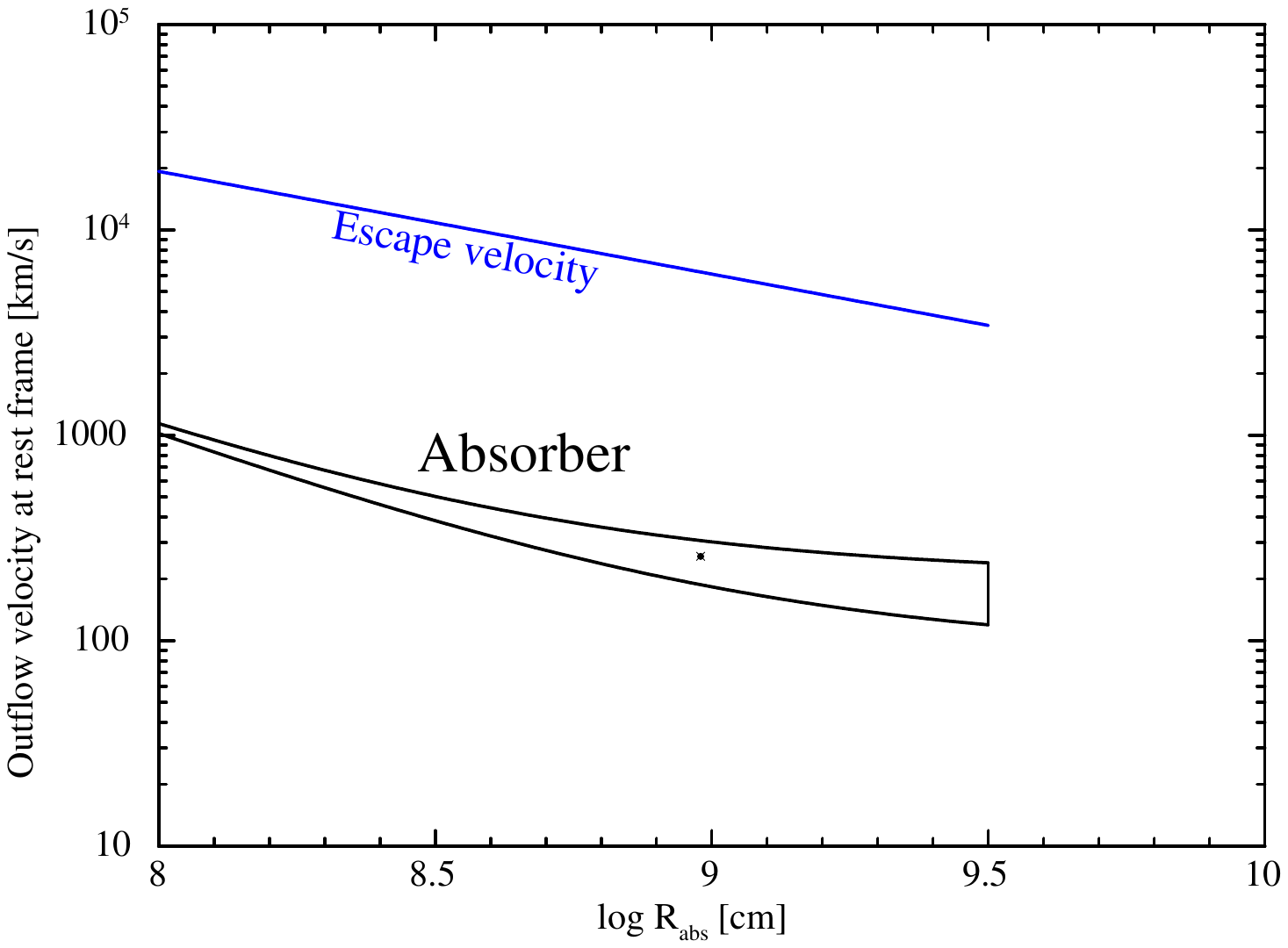}\\
  \end{center}
  \caption{%
    The outflow velocity in the rest frame, corrected for gravitational redshift and transverse Doppler effect. The cross mark indicates the best-fit value ( 260 km s$^{-1}$ ). The escape velocity is also plotted in blue.}
{Alt text: The horizontal axis represents \( R_{\rm abs} \). The error region obtained from the fitting is indicated by black lines.}

   \label{fig:correctedvelocity}
\end{figure}

\subsection{Comments on accretion disk and its X-ray luminosity }

Figure~\ref{fig:lightcurves} shows a nearly constant flux in the persistent phase, whereas \citet{Hyodo2009} observed a dip in the high/soft state of \axj. \citet{Ponti2015} and \citet{Ponti2017} also noted several occurrences of dips in the high/soft state. The absence of the dip in our observations suggests that the dip structure has moved away from the line of sight. This movement could be explained by disk precession, tidally driven by a companion star, as suggested in some low-mass X-ray binaries (e.g., XTE~J1118$+$480: \cite{Whitehurst1991, Torres2002}). Since our observation spans about 9 orbits, the lack of the dip implies that the precession period is at least an order of magnitude longer than the orbital period of $\sim30,000$~s.


The all extinction collected bolometric luminosity of $0.4 \times 10^{38} \ {\rm erg\ s^{-1}}$ corresponds to 0.1--0.2 times the Eddington luminosity. 
The accretion disk that achieves this luminosity is theoretically classified as a standard accretion disk (e.g., \cite{Shakura1973}). The comments given here may be applicable for a general understanding of the precession and the disk wind-driving mechanism in the standard accretion disk.



%

\section{Summary}

We observed the low-mass X-ray binary \axj\ using the \resolve\ and \xtend\ instruments onboard the X-ray astronomy satellite \xrism. The \resolve\ instrument provides excellent energy resolution, while \xtend\ offers broad energy coverage.
The source \axj\ was located just outside the field of view of \resolve, resulting in only a small number of photons being detected due to spillover from the mirror’s point spread function (PSF). In contrast, \xtend\ fully captured \axj\ within its field of view, enabling sufficient photon statistics.
Given these observational conditions, we performed data reduction in a manner that leveraged the respective strengths of each instrument and conducted the spectral analysis accordingly.

The \xtend\ light curve revealed the system's 8.4-hour eclipsing behavior. The spectra during the eclipse were mostly explained by interstellar dust scattering. The observed flux in the 3--6~keV band was \(7.1 \times 10^{-11}~\mathrm{erg~s^{-1}~cm^{-2}}\), identifying the system as being in a soft and high state. Interestingly, no clear dips were detected in the light curve, suggesting that \axj\ was in a dip-less soft/high state.

The spectrum obtained with \resolve\ clearly showed very narrow Fe and Ni absorption lines. However, the number of photons collected via PSF spillover was limited, and the complex background in this region posed challenges for its accurate subtraction from the source spectrum.
To mitigate these issues, we restricted the spectral fitting of the \resolve\ data to the narrow energy band of 6.8--9~keV, where the results are less sensitive to uncertainties in background subtraction.


A distinct separation of the Fe Ly$\alpha_1$ and $\alpha_2$ lines strongly proves a small line shift with a narrow width. In fact, the best-fit \texttt{ionabs}\ model shows that the blueshift of the absorption lines is as low as  \(160^{+50}_{-70} ~\mathrm{km~s^{-1}}\), and the broadening is as narrow as \(110^{+40}_{-30}~\mathrm{km~s^{-1}}\).   This narrow line width is consistent with binary orbital motion, suggesting that the absorber is moving with suppressed turbulence. 

Taking advantage of \xtend's broad energy sensitivity, we applied a photoionization plasma code \texttt{pion} to the spectrum. By using the eclipse-phase spectra as the background for the persistent emission, we achieved robust background subtraction. However, since the eclipse duration accounts for only about 5\% of the total observation time, this method was applicable only to the \xtend\ data, which provided sufficient photon statistics.

The broad-band \xtend\ spectra in the 2--12~keV range yielded a high ionization parameter of 

\(\log \xi = 4.4^{+0.3}_{-0.2}\), 
with a thick absorbing column density of \(N_{\rm H (pion)} = 16^{+6}_{-7} \times 10^{23}~\mathrm{cm^{-2}}\)  for a single-component photoionized absorber model. This suggests that the absorbers are located at \(R_{\rm abs} \lesssim 3.2 \times 10^{9}~\mathrm{cm}\), corresponding to \(1.6 \times 10^{4} \, R_{\rm g}\).  
The distance \(R_{\rm abs}\) is orders of magnitude smaller than the Roche lobe radius.
When combining the location-based constraints from the absorber in the simple single \texttt{pion} model measured by \xtend\ and the velocity shift measured by \resolve, it results in a solution, the outward velocity of the absorber is much slower than the escape velocity from the neutron star's gravity. 


Since the \resolve\ data taken in 2024 are limited in photon statistics, both for the source and the background, only a simplified model can be applied for the fittings. Further detailed modeling should be conducted with more photon statistics to reduce both statistical and systematic uncertainties on the parameters. Intense monitoring of \axj\ with \resolve\ will help achieve this.

\section*{Acknowledgement}

We would like to thank Kai Matsunaga, Maxime Parra, and Teruaki Enoto for their valuable comments and suggestions on the analysis of \axj.
Part of this work was support by JSPS KAKENHI grant numbers 23K22548, 23K20862, 24K00677, 21K03615, 24K17093, 23H00128, 24K00677,24K17105, and  24KJ0152 and NASA under award No. 80NSSC23K0738. 
This work was partly achieved through the use of SQUID at D3 Center, Osaka University.

\bibliographystyle{pasj} 
\bibliography{reference}   

\appendix


\appendix

\

\section{Raw spectra of \axj\ and its background obtained with \resolve}
\label{app:background}

In this Appendix, we evaluate the impact of background subtraction and associated systematic uncertainties on the absorption-line measurements presented in the main text. This material summarizes the detailed analysis performed to address concerns regarding the robustness of the Resolve spectra, particularly in the Fe~K band. It is provided here to document the technical limitations of the measurements while preserving the readability of the main scientific results.

\subsection{Raw Spectra and Background Components}


We extracted the source spectra from four pixels near \axj\ and the background spectra from four pixels indicated in Figure~\ref{fig:images}. These spectra are shown in Figure~\ref{fig:resolve_raw}.

An absorption feature corresponding to the Fe Ly$\alpha$ lines at 7.0 keV is clearly visible even in the raw source spectrum.
In contrast, the energy band around the Fe He-like K lines appears as a blend of multiple emission lines, heavily contaminated by background Fe He-like lines originating from the Galactic center diffuse emission and the nearby supernova remnant \sgraeast.

\begin{figure}[h]
  \begin{center}
  \begin{center}
\includegraphics[width=0.48\textwidth,angle=0]{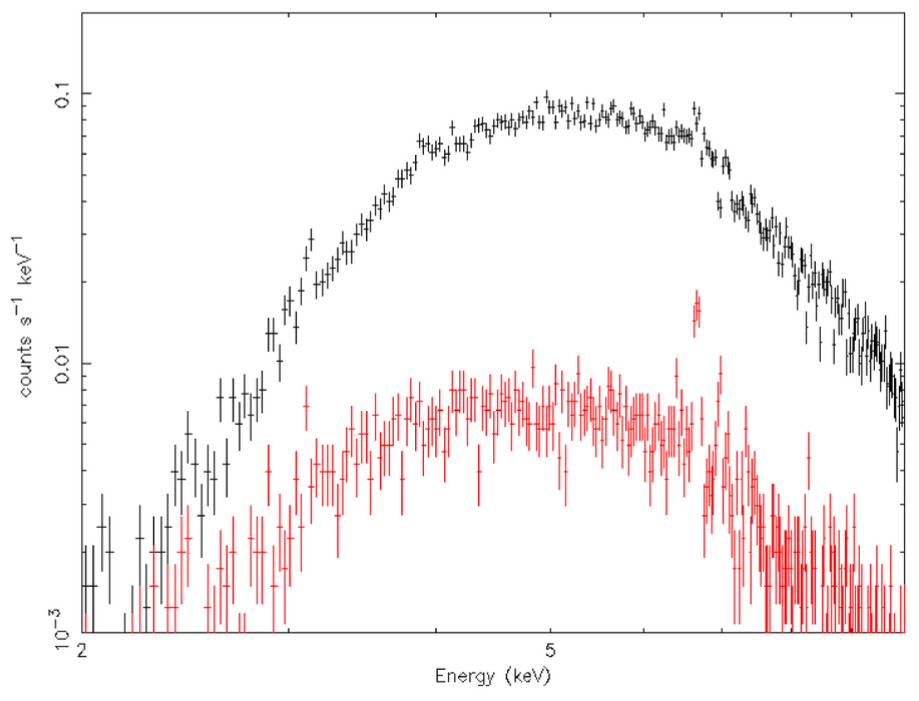}
    \end{center}
  \end{center}
  \caption{%
Black: \resolve\ source spectrum in the 2--10~keV band but without background subtraction. 
Red: \resolve\ background spectrum. The spectra were binned for clarity in display.
     }
      {
  Alt text:The horizontal axis represents energy, while the vertical axis shows intensity in units of \(\mathrm{c/s/keV}\).
  }
  \label{fig:resolve_raw}
\end{figure}


\subsection{Fe He-like Band}

Figure~\ref{fig:app_r2} presents the raw source spectra overlaid with preliminary models of the Galactic diffuse background (Uchiyama et al. in preparation) and the \sgraeast\ (\cite{Sgraeast2024}) contamination. \sgraeast\ is known to exhibit strong emission lines from He-like ions such as Si, S, Ar, and Fe. A major challenge in modeling its contribution to the \axj\ extraction region is the uncertainty in the effective-area (ARF) normalization in the PSF-tail region.

For the present dataset, the uncertainty in the PSF-tail ARF normalization is estimated to be approximately 50\% (\cite{Sgraeast2024}). 
When such uncertainties are applied to the Sgr~A~East component, it becomes difficult to determine whether apparent spectral features in the He-like band originate from genuine absorption or from residual emission contamination. For this reason, we excluded the Fe He-like band from the quantitative spectral modeling presented in the main text.

\subsection{Fe H-like Band}

The Fe H-like absorption lines are not completely immune to background subtraction; however, their sensitivity to background uncertainty is limited. Figure~\ref{fig:app_r3} shows the raw and background-unsubtracted spectra overplotted with the best-fit models derived from the background-subtracted spectra reported in Table~\ref{tb:ionabs} of the main text. In the raw spectra, the absorption depth reaches approximately 40\% of the continuum, whereas only $\sim$10\% remains after subtracting the adopted background. Taking the instrumental energy resolution into account, the line core is consistent with near-zero flux in both cases. Importantly, the two components of the Fe Ly$\alpha$ doublet (K$\alpha_1$ and K$\alpha_2$) are clearly resolved.


To quantify the impact of background subtraction on the derived line parameters, we examined three cases: (1) the standard background subtraction adopted in this paper, (2) no background subtraction (undersubtraction), and (3) subtraction with 1.5 times the adopted background level (oversubtraction). We modeled the spectra using the \texttt{ionabs} model, which simultaneously fits multiple transitions including Fe Ly$\alpha$, Ly$\beta$, and Ly$\gamma$. The results are summarized in Table~\ref{tab:app_r1}.

Even in the oversubtraction case, where the line bottom reaches zero count rate, the best-fit velocity shifts remain consistent within uncertainties. This demonstrates that the inferred line-of-sight velocities are robust against reasonable variations in background subtraction.

The analysis further shows that the line width and velocity shift are most tightly constrained in the no-subtraction case, where photon statistics are highest. In this case, the turbulent velocity is constrained to $V_{\rm turb} = 30 \pm 10$~km~s$^{-1}$. In the background-subtracted cases, the uncertainties increase substantially, indicating that the errors are dominated by photon statistics rather than by systematic effects. In all three cases, we obtain constraints of $V_{\rm turb} < 200$~km~s$^{-1}$ and $|V_{\rm los}| < 220$~km~s$^{-1}$, thereby excluding velocity components exceeding $\sim$200~km~s$^{-1}$. These results indicate that the absorption lines are intrinsically narrow and that the absorber exhibits only modest bulk and turbulent motions.





\begin{figure}
\centering
\includegraphics[width=\linewidth]{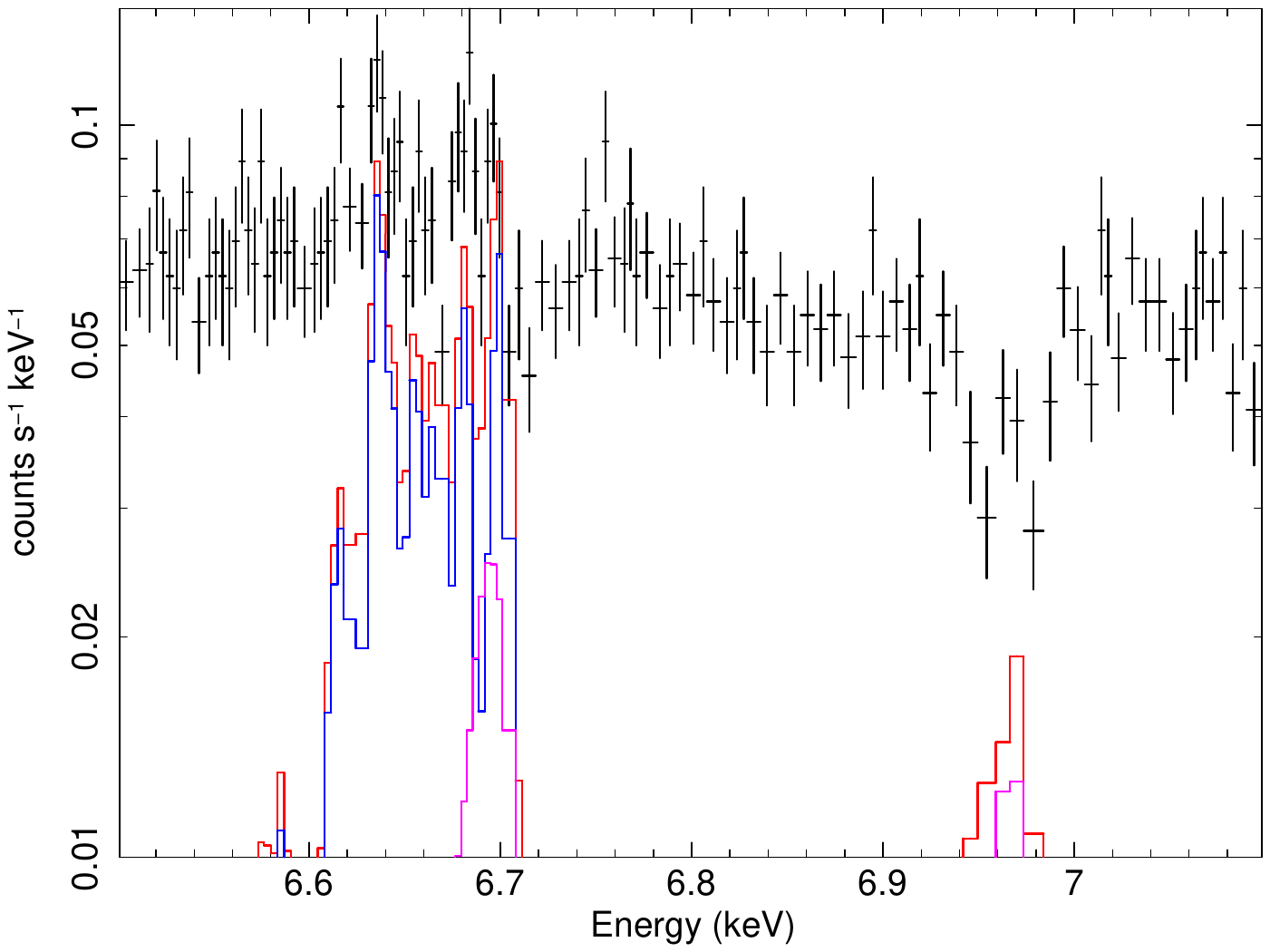}
\includegraphics[width=\linewidth]{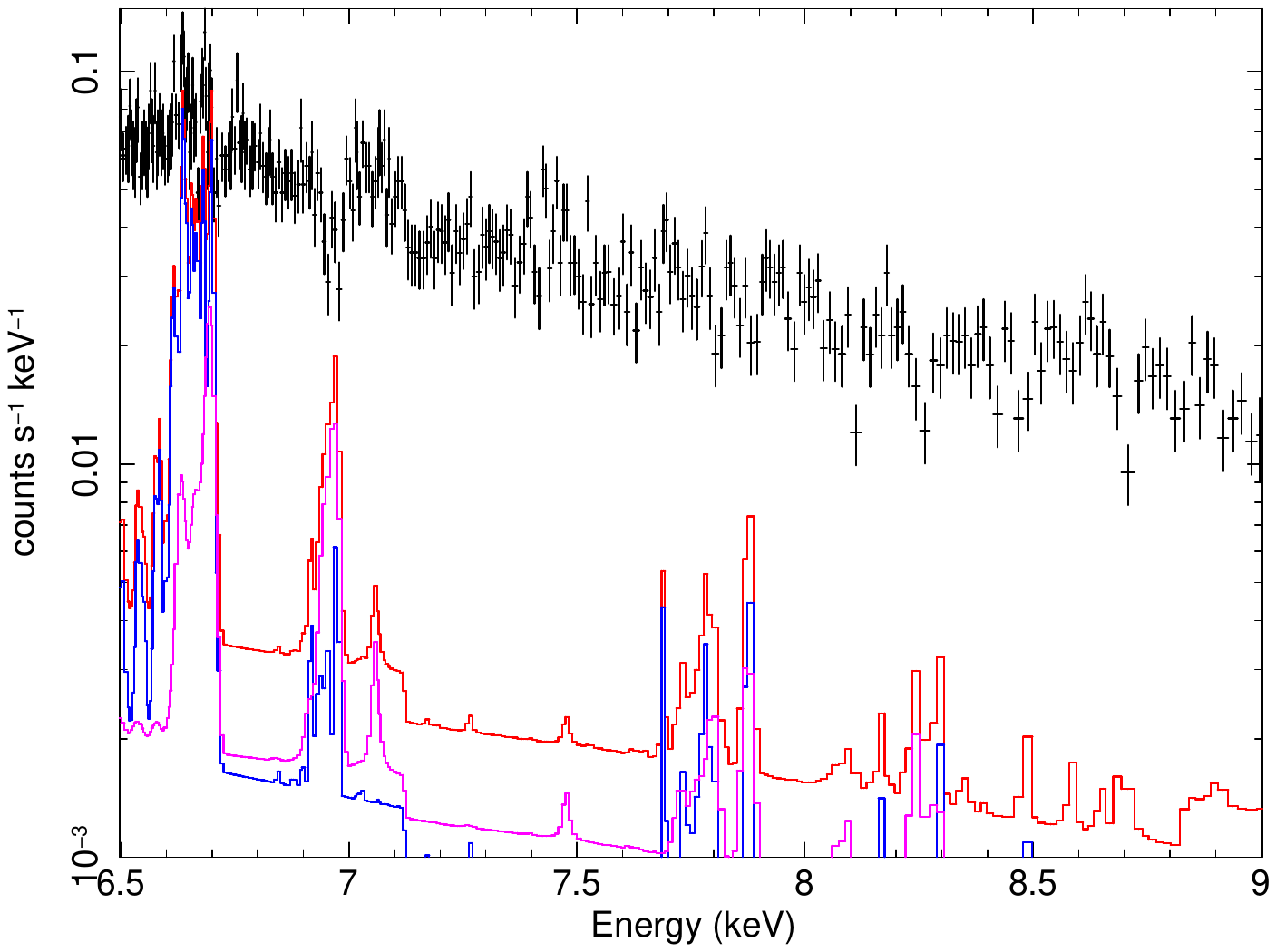}
\caption{
\ 
Raw Resolve source spectra overlaid with preliminary background models.
(upper) 6.5--7.1~keV band; (lower) 6.5--9~keV band.
The red curve shows the total background (Galactic diffuse emission plus Sgr~A~East contamination),
the blue curve shows the Sgr~A~East component,
and the magenta curve represents the Galactic diffuse emission.
The Galactic diffuse background spectrum is derived from offset observations (GC2 and GC3, Uchiyama et al. in prep.), located $\sim$10~arcmin away from \axj.
}
\label{fig:app_r2}
\end{figure}

\begin{figure}
\centering
\includegraphics[width=1.1\linewidth]{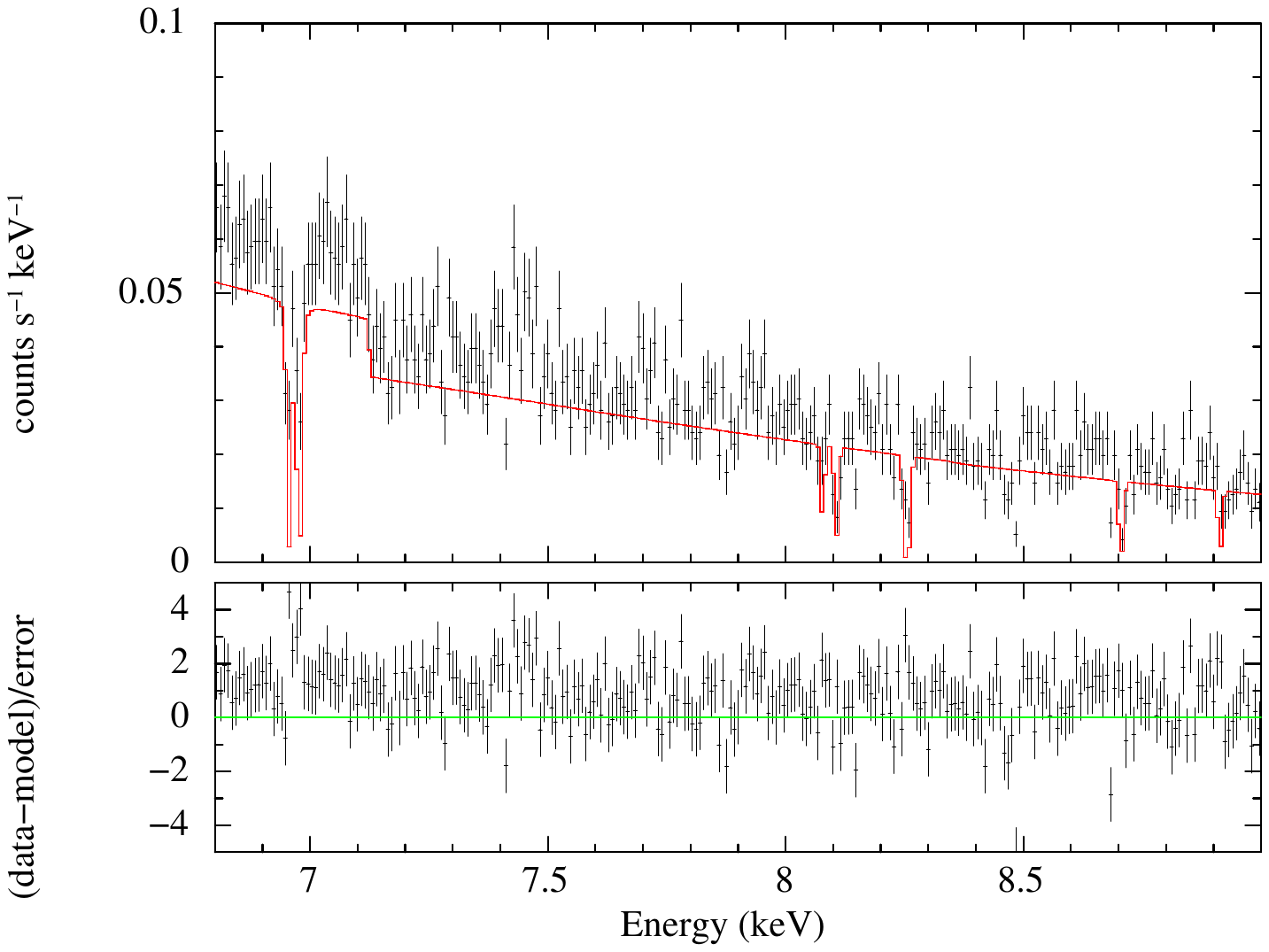}
\includegraphics[width=1.1\linewidth]{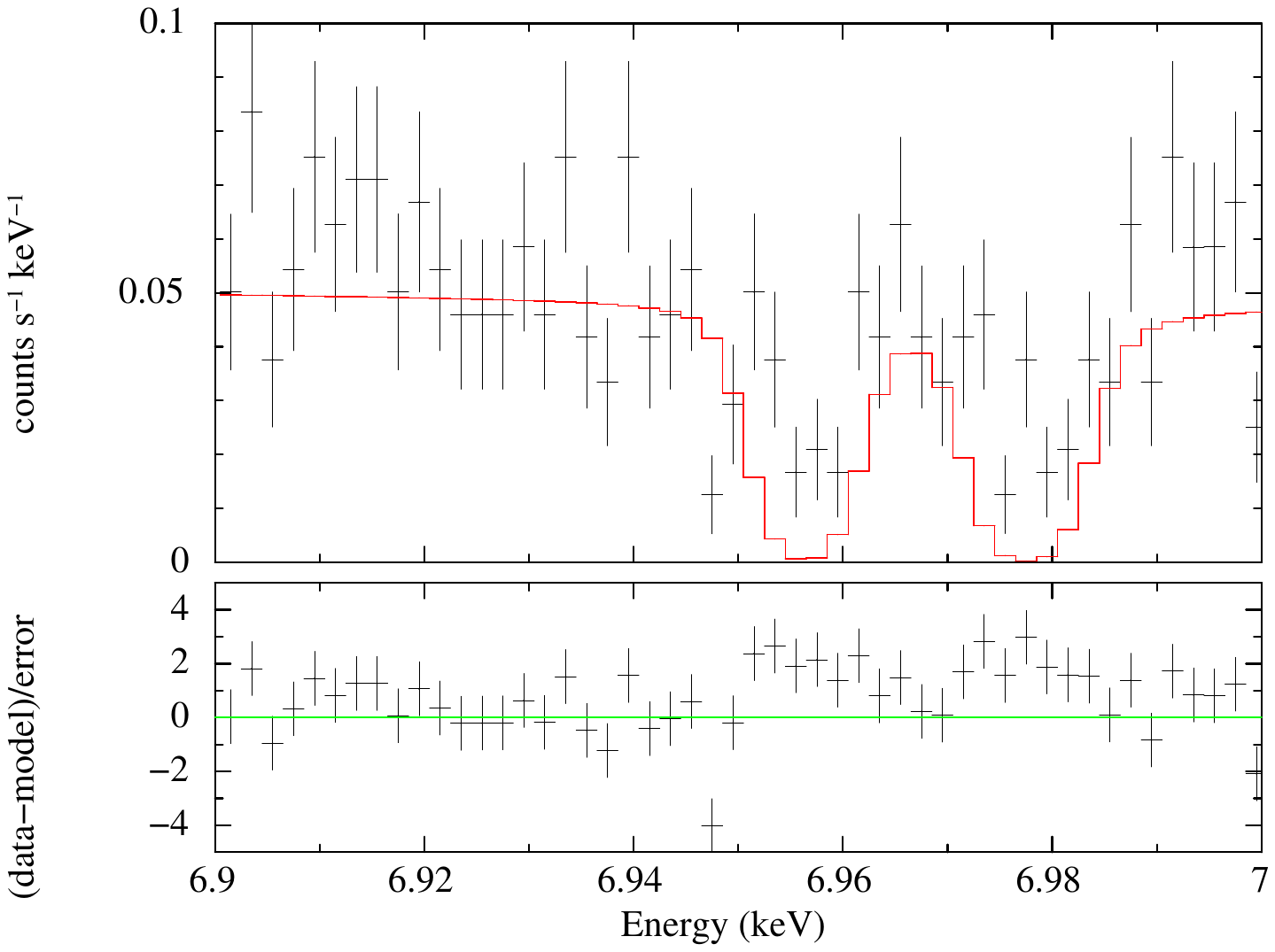}
\caption{
\ 
Raw Resolve source spectra in the H-like Fe K band.
(top) 6.8--8.0~keV; (bottom) 6.9--7.0~keV.
The red curve shows the best-fit model derived from the background-subtracted spectra reported in Table~1 of the main text (i.e., not a fit to the raw spectra).
}
\label{fig:app_r3}
\end{figure}

\begin{table}
\centering
\caption{\ Results of \texttt{ionabs} modeling of the Resolve spectra with different background treatments.}
\label{tab:app_r1}
\text{
\begin{tabular}{lccc}
\hline
Background case &
$V_{\rm turb}$  &
$V_{\rm los}$  &
$N_{\rm Fe}$  \\
 &
[km~s$^{-1}$] &
[km~s$^{-1}$]  &
[$10^{18}$~cm$^{-2}$] \\
\hline
(1) Standard subtraction &
$110_{-30}^{+40}$ &
$160_{-70}^{+50}$ &
$25_{-10}^{+30}$ \\
(2) No subtraction &
$30_{-10}^{+10}$ &
$190_{-30}^{+30}$ &
$12_{-6}^{+7}$ \\
(3) 1.5$\times$ background &
$140_{-30}^{+40}$ &
$80_{-70}^{+80}$ &
$60_{-30}^{+110}$ \\
\hline
\end{tabular}
}
\end{table}




\subsection{Photo-ionization modeling including the Fe He-like Band}

As we made for the \xtend\ spectra in section \ref{sec:xtendpion}, we also applied the \texttt{pion} model to the \resolve\ spectra, but with an energy range of 5--10~keV including the Fe He-like band. 
\
For this fitting, we used the C-statistic. The best-fit parameters are shown in Figure~\ref{fig:resolvepion} and listed in Table~\ref{tb:Pion_rsl}. The ionization parameter $\log \xi$ was  $4.3^{+0.3}_{-0.1}$,   consistent with the value obtained using \xtend. The column density was  $N_{\rm H (pion)} = 17^{+15}_{-5} \times 10^{23}$~cm$^{-2}$,   which is also consistent with the \xtend\ results.

While the best-fit parameters from \resolve\ were generally consistent with those from \xtend, large residuals around the \Fevc\ line at 6.7~keV were observed, likely due to an insufficient subtraction of the contamination from \sgraeast\ and the Galactic diffuse emission. 
Therefore, throughout this paper, we adopted the best-fit parameters of the photo-ionization model derived from the \xtend\ data that provide the robust background subtraction.

A deep absorption line corresponding to the resonance transition of He-like iron is likely present around 6.700 keV in the rest frame, with a narrow width comparable to that of the Fe Ly$\alpha$ absorption lines.
In addition, the associated forbidden line at 6.637 keV and intercombination lines at 6.682 keV and 6.668 keV were marginally identified; however, their statistical significance was insufficient, and they did not meet the accuracy requirements of the spectral fitting performed in this study.

\begin{figure}[h]
  \begin{center}
    \includegraphics[width=0.48\textwidth,angle=0]{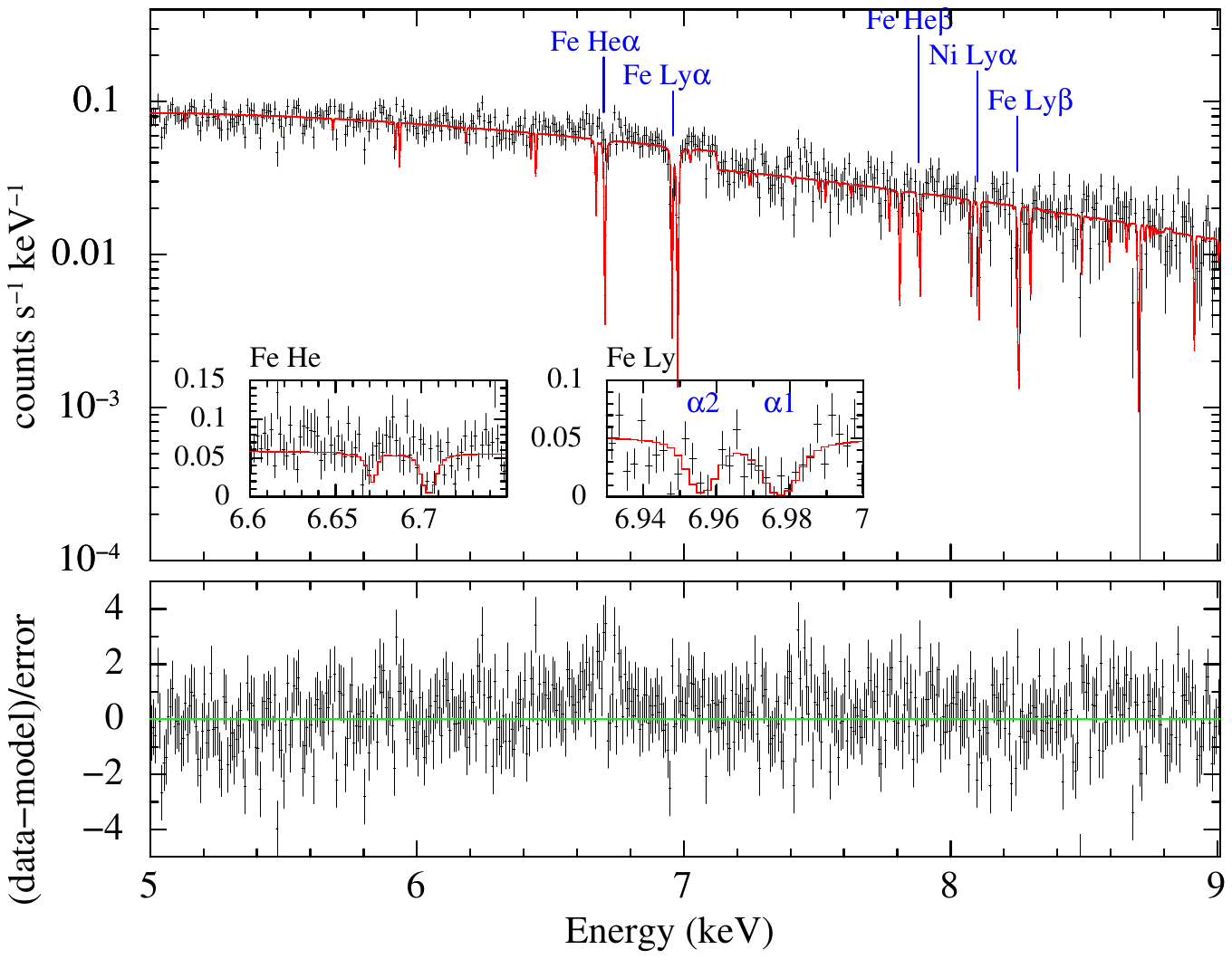} \\
  \end{center}
  \caption{
\resolve\ 5--10 keV band spectra with the \texttt{pion} model. The spectra were binned for clarity in display.
 }
   {
  Alt text: The horizontal axis represents energy, while the vertical axis shows intensity in units of \(\mathrm{c/s/keV}\) in the upper panel, and residuals in units of \((\mathrm{data} - \mathrm{model})/\mathrm{error}\) in the lower panel. The data and residuals are shown in black, while the model is shown in red.}
   \label{fig:resolvepion}
\end{figure}

\begin{table*}[tbph]
 \begin{center}
  \caption{\texttt{Pion} models for absorption lines above 5.0 keV (\resolve)}
 \label{tb:Pion_rsl}
 \begin{tabular}{lccc} 
 \hline \hline
\hline
  \multicolumn{4}{l}{Model : TBabs $\times$ \texttt{pion} $\times$  (Disk blackbody $+$ Blackbody) } \\ \hline
    \multicolumn{4}{l}{Interstellar absorption {\it TBabs}..........} \\
 & $N_{\rm H \ (TBabs)}$ & [$10^{23}$cm$^{-2}$] & 1.70(fix)  \\
\multicolumn{4}{l}{Absorption by photoionized materials  \texttt{pion} ..........} \\
  & log\(\xi\) & [erg cm s$^{-1}$] &  $4.3^{+0.3}_{-0.1}$  \\
   & $N_{\rm H (pion)}$   & [$10^{23}$cm$^{-2}$] &   $17^{+15}_{-5}$ \\
    & $V_{\rm rms\ \rm(pion) }$  & [km s$^{-1}$] &   $30^{+50}_{-10}$ \\
          & $z_{\rm \ (pion)}$  &  &  $-5.2^{+0.8}_{-2.0}\times10^{-4}$$^*$ \\
        \multicolumn{4}{l}{Disk blackbody ..........} \\
 & $kT_{\rm in\ {(DiskBB)}}$  & [keV] & 1.14 (fix) \\
 & Normalization$_{\rm \ (DiskBB)}$ & [$(R_{\rm in\ km}/D_{\rm 10\ kpc})^2 \cos\theta$] &   90 (fix) \\
    \multicolumn{4}{l}{Blackbody ..........} \\
 & $kT_{\rm (BB)}$  & [keV] & 2.0(fix) \\
& Normalization$_{\rm \ (BB)}$ & [$L_{39\ \rm erg\ s^{-1}}/D_{\rm 10\ kpc}^2$]  &   1.4$\times10^{-2}$ (fix) \\ \hline
  \multicolumn{4}{r}{Cstatics   9728.77 using 9999 bins} \\ \hline
\end{tabular}
\end{center}
We fixed the continuum model, {\it TBabs} $\times$ ({\it Disk blackbody} $+$ {\it Blackbody}), to the best-fit parameters determined by the \xtend's \texttt{pion} fitting.
$^*$ The correction of the energy shift caused by Earth's orbital motion around the Sun ($-28 \, \text{km s}^{-1}$) was applied. The $z_{\rm \ (pion)}$ before the correction is   $-6.0\times10^{-4}$. 
\end{table*}

\section{Raw persistent and eclipse spectra of \xtend}

To make the background-subtracted spectra shown in Figure~\ref{fig:xtendpion}, we extracted the persistent spectra (outside of eclipse) and the background spectra (during eclipse) from a circular region with a radius of $1\farcm$ centered on \axj. 
For comparison, we also extracted the persistent and background spectra from a larger circular region with a radius of $5\farcm$, also centered on \axj. Both sets of spectra are presented in Figure~\ref{fig:Xtend_raw_bgd2}. The results of spectral fitting using the \texttt{pion} model are summarized in Table~\ref{tb:Pion}. The prominent He-like Fe emission line seen in the $5\farcm$ spectra is primarily attributed to the supernova remnant Sgr~A~East, which lies within this extraction region.

\begin{figure}[h]
  \begin{center}
  \begin{center}
\includegraphics[width=0.48\textwidth,angle=0]{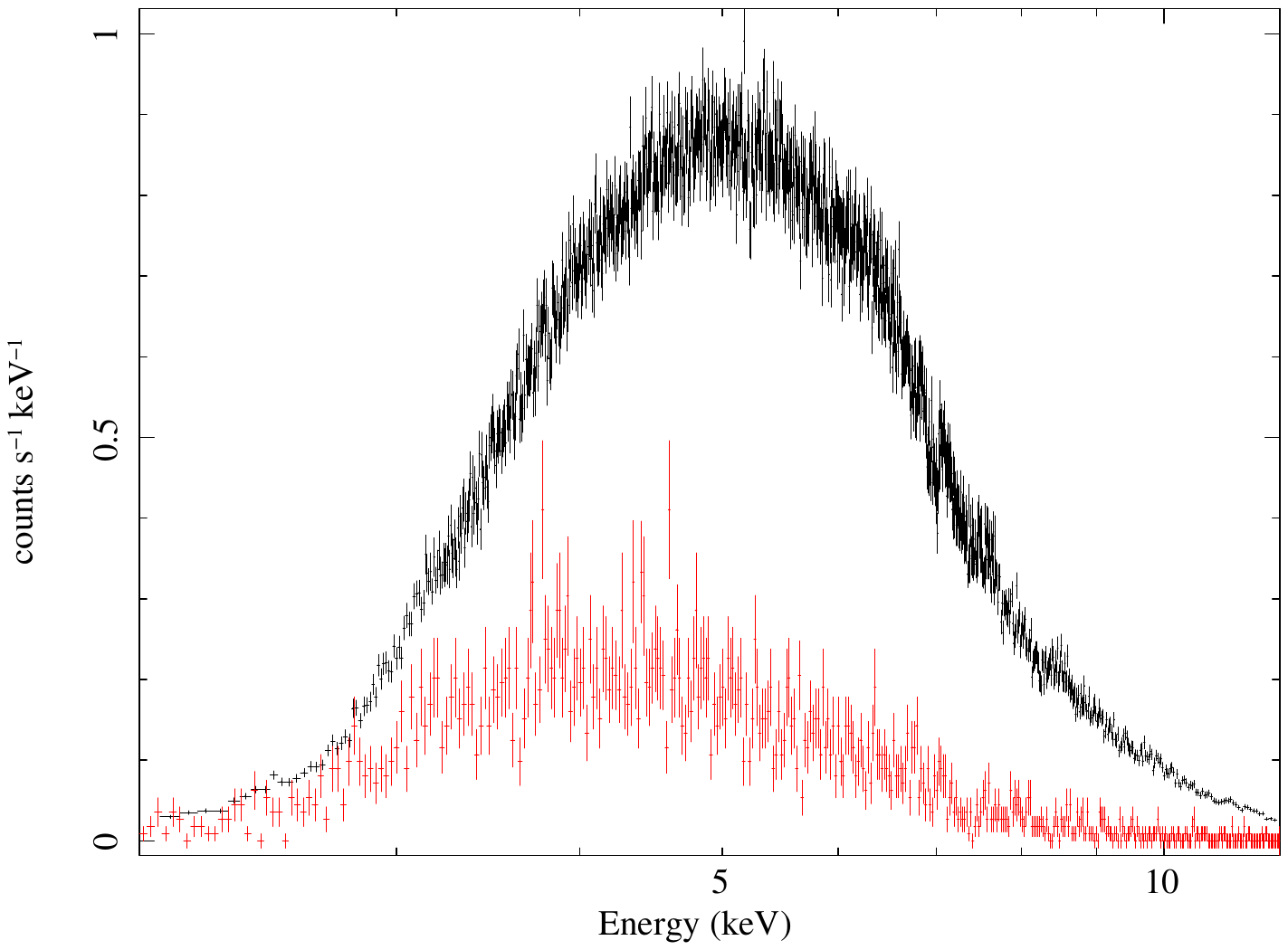}
\includegraphics[width=0.48\textwidth,angle=0]{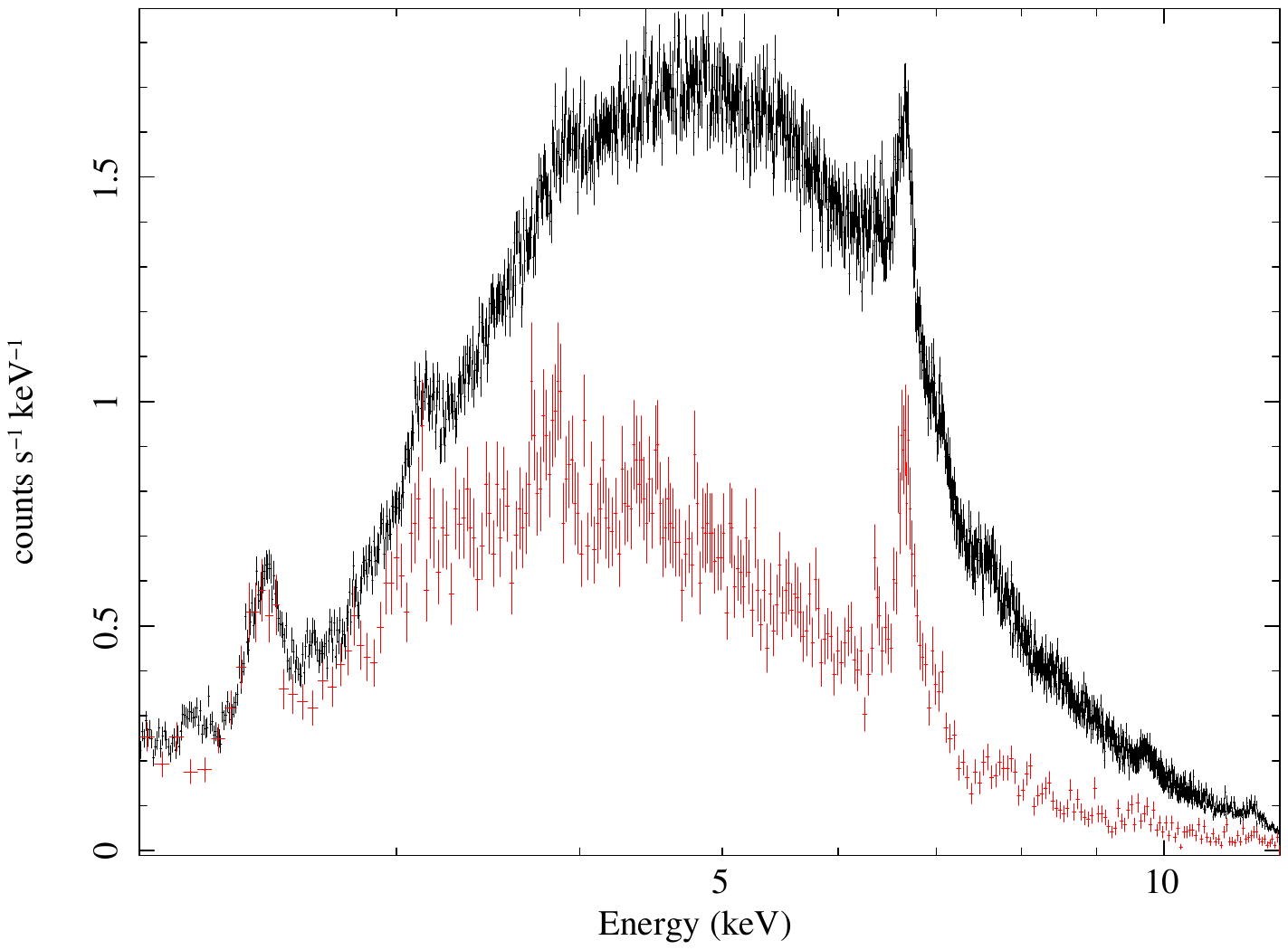}
    \end{center}
  \end{center}
  \caption{
\xtend\ raw spectra without background subtraction. 
Upper:source spectrum outside of eclipse (black) and during the eclipse (red) extracted from the $1\farcm$ circular region. Bottom: Same but from the $5\farcm$ circular region. The spectra were binned for clarity in display.}
{
  Alt text:The horizontal axis represents energy, while the vertical axis shows intensity in units of \(\mathrm{c/s/keV}\).
  }
  \label{fig:Xtend_raw_bgd2}
\end{figure}

\end{document}